\documentclass[fleqn,usenatbib]{mnras}

\usepackage{newtxtext,newtxmath}

\usepackage[T1]{fontenc}
\usepackage{ae,aecompl}

\usepackage{comment}
\usepackage{pgfplotstable}
\usepackage{csvsimple}
\usepackage{booktabs} 
\usepackage{hyperref}
\usepackage{soul}

\usepackage[version=4]{mhchem}

\usepackage{graphicx}	
\usepackage{amsmath}	

\newcommand{\kms}{km\,s$^{-1}$}
\newcommand{\Ha}{H$\alpha$}
\newcommand{\Paa}{Pa$\alpha$}
\newcommand{\Bra}{Br$\alpha$}
\newcommand{\Brg}{Br$\gamma$}

\newcommand{\HI}{\ion{H}{i}}
\newcommand{\HeI}{\ion{He}{i}}

\newcommand{\NII}{[\ion{N}{ii}]}

\newcommand{\FeI}{[\ion{Fe}{i}]}
\newcommand{\SiI}{[\ion{Si}{i}]}
\newcommand{\FeII}{[\ion{Fe}{ii}]}

\newcommand{\Fi}{$I_{\rm \nu\, F150W}$}
\newcommand{\Fii}{$I_{\rm \nu\, F164N}$}
\newcommand{\Fiii}{$I_{\rm \nu\, F200W}$}
\newcommand{\Fiv}{$I_{\rm \nu\, F212N}$}
\newcommand{\Fvi}{$I_{\rm \nu\, F356W}$}
\newcommand{\Fvii}{$I_{\rm \nu\, F405N}$}
\newcommand{\Fviii}{$I_{\rm \nu\, F444W}$}

\title[JWST/NIRCam imaging of SN\,1987A]{Deep JWST/NIRCam imaging of Supernova\,1987A}

\author[M. Matsuura et al.]{
Mikako Matsuura,$^{1}$ \thanks{E-mail: matsuuram@cardiff.ac.uk (MM)}
M. Boyer,$^{2}$ 
Richard G. Arendt,$^{3, 4}$ 
J. Larsson,$^{5}$ 
C. Fransson,$^{6}$ 
A. Rest,$^{2, 7}$ 
A. P. Ravi,$^{8}$ 
            \newauthor 
S. Park,$^{8}$ 
P. Cigan,$^{9}$ 
T. Temim,$^{10}$ 
E. Dwek,$^{3}$ 
M.J. Barlow,$^{11}$ 
P. Bouchet,$^{12,13}$ 
G. Clayton,$^{14}$ 
R. Chevalier,$^{15}$ 
           \newauthor 
J. Danziger,$^{16}$ 
J. De Buizer,$^{17}$ 
I. De Looze,$^{18}$ 
G. De Marchi,$^{19}$
O. Fox,$^{2}$
C. Gall,$^{20}$
R. D. Gehrz,$^{21}$
           \newauthor 
H. L. Gomez, $^{1}$
R. Indebetouw,$^{15, 22}$ 
T. Kangas,$^{23,24}$ 
F. Kirchschlager,$^{11, 18}$ 
R. Kirshner,$^{25}$ 
P. Lundqvist,$^{5}$ \newauthor
J.M. Marcaide,$^{26}$ 
I. Mart\'i-Vidal,$^{26, 27}$ 
M. Meixner,$^{28}$
D. Milisavljevic,$^{29,30}$
S. Orlando,$^{31}$
M. Otsuka,$^{32}$  \newauthor
F. Priestley,$^{1}$
A.M.S. Richards,$^{33}$
F. Schmidt, $^{11}$ 
L. Staveley-Smith,$^{34}$ 
Nathan Smith,$^{35}$         
J. Spyromilio,$^{36}$ \newauthor
J. Vink, $^{37, 38, 39}$ 
Lifan Wang,$^{40}$ 
D. Watson,$^{41, 42}$
R. Wesson,$^{1, 11}$
J. C. Wheeler,$^{43}$ 
C.E. Woodward,$^{21}$  
G. Zanardo,$^{33}$   \newauthor
D. Alp,$^{5}$     
D. Burrows,$^{44}$
\\
$^{1}$ Cardiff Hub for Astrophysical Research and Technology (CHART), School of Physics and Astronomy, 
Cardiff University, The Parade, Cardiff CF24 3AA,  UK \\
$^{2}$ Space Telescope Science Institute, 3700 San Martin Drive, Baltimore, MD 21218, USA \\
$^{3}$ Observational Cosmology Lab, Code 665, NASA Goddard Space Flight Center, Greenbelt, MD 20771, USA \\
$^{4}$ University of Maryland, Baltimore County, Baltimore, MD 21250, USA \\
$^{5}$ Department of Physics, KTH Royal Institute of Technology, The Oskar Klein Centre, AlbaNova, Stockholm, SE-106 91, Sweden \\
$^{6}$ Department of Astronomy, Stockholm University, The Oskar Klein Centre, AlbaNova, Stockholm, SE-106 91, Sweden \\
$^{7}$ Department of Physics and Astronomy, The Johns Hopkins University, 366 Bloomberg Center,
3400 N. Charles Street, Baltimore, MD 21218, USA \\
$^{8}$ Box 19059, Department of Physics, University of Texas at Arlington, Arlington, TX 76019, USA \\
$^{9}$ U.S. Naval Observatory, 3450 Massachusetts Ave NW, Washington, 20392-5420, DC, USA \\
$^{10}$ Department of Astrophysical Sciences, Princeton University, Princeton, NJ 08544, USA \\
$^{11}$ Department of Physics and Astronomy, University College London (UCL), Gower Street, London WC1E 6BT, UK \\
$^{12}$ DRF/IRFU/DAp, CEA-Saclay, F-91191 Gif-sur-Yvette, France \\
$^{13}$ NRS/AIM, Universit\'e Paris Diderot, F-9119, Gif-sur-Yvette, France \\
$^{14}$ Department of Physics \& Astronomy, Louisiana State University, Baton Rouge, LA 70803, USA \\
$^{15}$ Department of Astronomy, University of Virginia, 530 McCormick Road, Charlottesville, VA 22904, USA \\
$^{16}$ Osservatorio Astronomico di Trieste, Via Tiepolo 11, Trieste, Italy \\
$^{17}$ SOFIA-USRA, NASA Ames Research Center, Mail Stop 232-12, Moffett Field, CA 94035, USA \\
$^{18}$ Sterrenkundig Observatorium, University of Ghent, Krijgslaan 281 - S9, B-9000 Ghent, Belgium \\
$^{19}$ European Space Research and Technology Centre, Keplerlaan 1, 2200 AG, Noordwijk, The Netherlands \\
$^{20}$ DARK, Niels Bohr Institute, University of Copenhagen, Jagtvej 128, Copenhagen, 2200, Denmark \\
$^{21}$ Minnesota Institute for Astrophysics, University of Minnesota, 116 Church Street S. E., Minneapolis, 55455, MN, USA \\
$^{22}$ National Radio Astronomy Observatory, 520 Edgemont Road, Charlottesville, 22903, VA, USA \\
$^{23}$ Finnish Centre for Astronomy with ESO (FINCA), University of Turku, 20014 Turku, Finland \\
$^{24}$ Tuorla Observatory, Department of Physics and Astronomy, University of Turku, 20014 Turku, Finland \\
$^{25}$ TMT International Observatory, 100 West Walnut Street
Pasadena CA 91124 USA, USA \\
$^{26}$ Departamento de Astronom\'ia y Astrof\'isica, Universidad de Val\`encia, C/Dr. Moliner 50, E-46100 Burjassot, Spain \\
$^{27}$  Observatori Astron\`omic, Universitat de Val\`encia, C. Catedr\'atico Jos\`e Beltr\'an 2, E-46980 Paterna, Val\`encia, Spain \\
$^{28}$ Jet Propulsion Laboratory, California Institute of Technology, 4800 Oak Grove Dr., Pasadena, CA 91109, USA\\
$^{29}$ Department of Physics and Astronomy, Purdue University, 525 Northwestern Avenue, West Lafayette, IN 47907, USA\\
$^{30}$ Integrative Data Science Initiative, Purdue University, WestLafayette, IN 47907, USA \\
$^{31}$ Istituto Nazionale di Astrofisica-Osservatorio Astronomico di Palermo, Piazza del Parlamento 1, 90134 Palermo, Italy \\
$^{32}$ Okayama Observatory, Organization, Street, City, 610101, State, Japan \\
$^{33}$ JBCA, School of Physics and Astronomy, University of Manchester, Manchester M13 9PL, UK \\
$^{34}$ International Centre for Radio Astronomy Research (ICRAR), The University of Western Australia, 35 Stirling Hwy, Crawley, WA 6009, Australia \\
$^{35}$ Steward Observatory, University of Arizona, 933 N. Cherry Ave., Tucson, AZ 85721, USA \\
$^{36}$ ESO, Karl-Schwarzschild-Str 2, Garching, 85748, Germany \\
$^{37}$ Anton Pannekoek Institute for Astronomy, University of Amsterdam, Science Park 904, NL-1098 XH Amsterdam, the Netherlands \\
$^{38}$ GRAPPA, University of Amsterdam, Science Park 904, NL-1098 XH Amsterdam, the Netherlands \\
$^{39}$ SRON, Netherlands Institute for Space Research, Sorbonnelaan 2, 3584 CA, Utrecht, the Netherlands \\
$^{40}$ Department of Physics and Astronomy, Texas A\&M University, College Station, TX 77843, USA \\
$^{41}$ Cosmic Dawn Center (DAWN), R\aa dmandsgade 62, 2200 K\o benhavn, Denmark \\
$^{42}$ Niels Bohr Institute, University of Copenhagen, Jagtvej 128, Copenhagen, 2200, Denmark \\
$^{43}$ Department of Astronomy, The University of Texas at Austin, 2515 Speedway, Austin, 78712-1205, TX, USA \\
$^{44}$ Dept. of Astronomy \& Astrophysics, Penn State University, University Park, PA 16802, USA \\
}

\date{Accepted 15th April 2024. Received 15th April 2024; in original form 1st September 2023}

\pubyear{2024}

\hypersetup{psdextra} 
\pgfplotsset{compat=1.18} 
\begin{document}
\label{firstpage}
\pagerange{\pageref{firstpage}--\pageref{lastpage}}
\maketitle

\clearpage
\begin{abstract}{
 {\it JWST}/NIRCam obtained high angular-resolution (0.05--0.1''), deep near-infrared 1--5\,\micron\, imaging of Supernova (SN) 1987A taken 35 years after the explosion.
In the NIRCam images, we identify: 1) faint H$_2$ crescents, which are emissions located between the
ejecta and the equatorial ring, 2) a bar, which is a substructure of the ejecta, and 3) the bright 3--5\,\micron\, continuum emission exterior
to the equatorial ring.
The emission of the remnant in the NIRCam 1--2.3\,\micron\, images is mostly due to line emission, which is mostly emitted in the ejecta and in the hot spots within the equatorial ring.
In contrast, the NIRCam 3--5\,\micron\, images are dominated by continuum emission. In the ejecta, the continuum is due to dust, obscuring the centre of the ejecta. In contrast, in the ring and exterior to the ring, synchrotron emission contributes a substantial fraction to the continuum. 
 Dust emission contributes to the continuum at outer spots and diffuse emission exterior to the ring, but little within the ring. 
This shows that dust cooling and destruction time scales are shorter than the synchrotron cooling time scale, and the time scale of hydrogen recombination in the ring is even longer than the synchrotron cooling time scale.
 With the advent of high sensitivity and high angular resolution images provided by {\it JWST}/NIRCam, our observations of SN\,1987A demonstrate that NIRCam opens up a window to study particle-acceleration and shock physics in unprecedented details, probed by near-infrared synchrotron emission,  building a precise picture of how a SN evolves.
 }
\end{abstract}



\section{Introduction}


The explosion of Supernova (SN) 1987A was detected on 23rd Feb 1987\footnote{\url{http://www.cbat.eps.harvard.edu/iauc/04300/04316.html}} in a neighbouring galaxy, the Large Magellanic Cloud, at a distance of $49.6$~kpc \citep{2019Natur.567..200P}. This is the nearest SN explosion detected in 400 years, since Kepler's SN  in 1604 \citep{McCray:1993p29839}.
Due to its proximity, SN\,1987A can be well-resolved with modern telescopes, and was observed at almost every wavelength from the $\gamma$-ray to the radio \citep{Arnett:1989p29666}.

\begin{figure*}
	 \includegraphics[width=16cm, trim={0cm 0cm 0cm 0cm},clip]{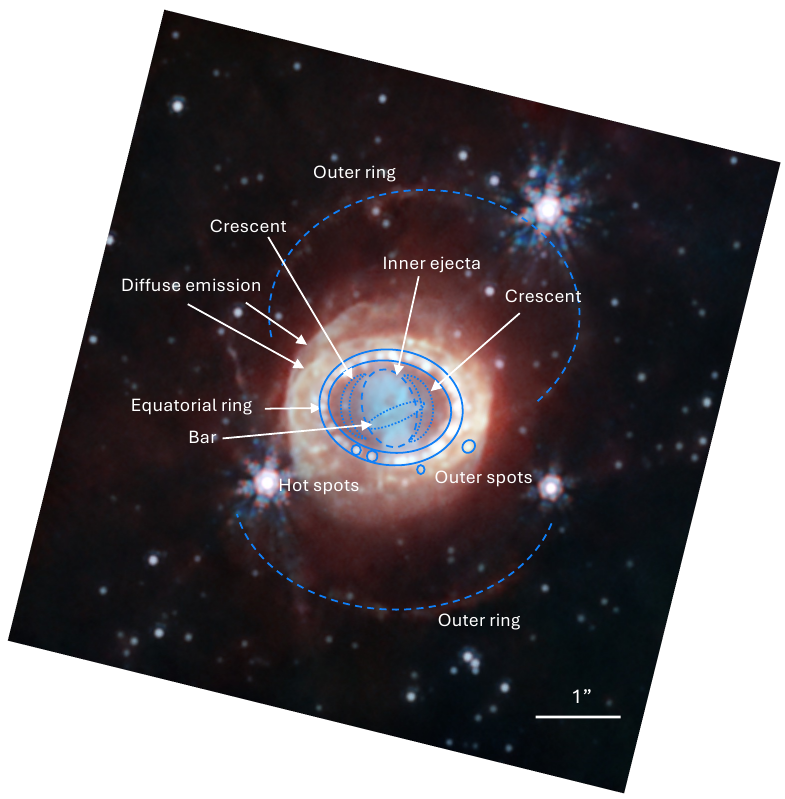}
        \caption{NIRCam five-colour image (Blue: F150W, Cyan: F164N, Cyan: F200W, Yellow: F323N, Orange: F405N, and Red: F444W), capturing the bright equatorial ring, shocked gas beyond the equatorial ring, and the very faint outer rings. 
        The inner ejecta stands out in cyan as it has strong F164N emission. 
        Within the inner ejecta, a bar crosses across approximately from the east to west.
        Two crescents are found in very faint blue/cyan emission between the inner ejecta and the equatorial ring. The hot spots are bright spots in the NIRCam/SW images within the equatorial ring, while the outer spots are found exterior of the equatorial ring.
         North is towards the top and east is to the left. 
        The original image was created by Alyssa Pagan (STScI), and the image credits belong to NASA, ESA, CSA, Mikako Matsuura (Cardiff University), Richard Arendt (NASA-GSFC, UMBC), Claes Fransson (Stockholm University), Josefin Larsson (KTH).
        \label{fig:three-color}}
\end{figure*}


The iconic {\it Hubble Space Telescope (HST)} optical image of SN 1987A consists of the inner ejecta, the equatorial ring, and the outer rings.
These components are also illustrated by our image (Fig.~\ref{fig:three-color})\footnote{The image is modified from the original image published at \url{https://webbtelescope.org/contents/media/images/2023/136/01H8Q02S452MC9CAF0VSJ3ZTFX }} , taken by NIRCam \citep{Rieke2023}
on board the {\it James Webb Space Telescope (JWST)} \citep{Gardner2023}.
The equatorial ring and outer rings are circumstellar material that was ejected from the progenitor \citep{1991ApJ...369L..63J, Crotts.1995}. It is not known exactly how these rings were formed, but most likely the process involved a binary system \citep[e.g.][]{2003LNP...598..219M}. 
It has been suggested that until about 20,000 years ago, the progenitor star was a red supergiant \citep{1991Natur.350..683C}.
The mass loss wind from the red supergiant was diverted to an equator by a binary system, and this process may eventually have led to the formation of the iconic rings around SN 1987A. \citep{1995MNRAS.273L..19L}.  \citet{Morris.2007} further proposed that the merger of a binary companion spun out the rings, and at the same time the merger turned the progenitor into a blue supergiant.
However, it still remains unknown when and how the progenitor turned into a blue supergiant, but it could involve dredge-up or binary merger \citep{1988Natur.334..508S,Arnett:1989p29666, 1989ApJ...336..429F}.
 A few Galactic examples of similar rings are seen around blue supergiants and luminous blue variables \citep{1997ApJ...489L.153B,2007AJ....133.1034S,2007AJ....134..846S,2013MNRAS.429.1324S}.

The outer ejecta expand with a speed exceeding 10,000\,km~s$^{-1}$ \citep{1988AJ.....95.1087P, McCray:1993p29839}, triggering shocks within the circumstellar material. 
Shocks brightened the equatorial ring (labelled in Fig.\,\ref{fig:three-color}) in about 1995 \citep{Sonneborn:1998p29919,2000ApJ...537L.123L}, and the brightness increased in X-ray, optical, infrared (IR) and radio until about 2010 \citep{Fransson:2015gp, Dwek:2010kv, Frank:2016ka}. 
Approximately in 2010, the shocks passed the equatorial ring, and material outside of the equatorial ring is now detected as faint outer spots (labelled in Fig.\,\ref{fig:three-color}) in the optical images \citep{Fransson:2015gp, Larsson.2019h9}.
Around this time, the light curves in different bands started to diverge. The hard X-ray and radio brightness continued to increase \citep{Frank:2016ka, Cendes.2018z98}, though at a slower rate than before 2003. 
Meanwhile, the near-infrared (NIR) brightness, as observed using {\it Spitzer}, peaked around 2010 and started declining thereafter \citep{Arendt:2016ds, Arendt2020}. 
The cause of diverged light curves at different wavelengths was unknown. 
The mid-IR emission is known to be from silicate dust in the equatorial ring \citep{Bouchet:2006p2168, Dwek:2010kv, Arendt:2016ds, Matsuura.2022fk}.
Although it has been suggested to be `hot' ($\sim$500\,K) dust \citep{Bouchet:2006p2168, Dwek:2010kv},
the emitting region and the source of the NIR 3--4\,$\mu$m flux was not well established,
because no instrument had the resolution and sensitivity to clearly spatially resolve SN\,1987A at 3--4\,$\mu$m before {\it JWST}. 
Modelling the total flux of near- to mid-IR spectra from {\it JWST}/NIRSpec and MIRI shows that the combination of dust, synchrotron, and atomic emission contributes to overall emission at this wavelengths \citep{Larsson.2023ytc, Jones.20232sn}. 
NIRCam imaging provides the highest angular resolution and best sensitivity available with {\it JWST}, and can pinpoint the emitting regions of different components of dust, synchrotron and atomic lines \citep{2023arXiv230913011A}. 
This will allow us to understand precisely the physics of dust, synchrotron, and atomic line emission in shocked and post-shocked regions in SN\,1987A, enabling us to witness the real-time physics of an evolving SN.

\section{JWST/NIRCam observations of SN 1987A} 
\label{sect-observations}

We obtained deep images of SN 1987A, using NIRCam on {\it JWST}  under general observer program 1726.
The data were acquired on 2022 September 1 and 2, days 12,974 and 12,975 after the SN explosion was first detected.
NIRCam has a short wavelength channel (0.6--2.3\,$\mu$m, hereafter SW) and a long-wavelength channel (2.4--5.0\,$\mu$m, hereafter LW).
Eight filters were used: four in the SW channel,  and four in the LW channel.
The properties of the filters are summarised in Table\,\ref{observing_log}.

\begin{figure}
	\includegraphics[width=8.7cm, trim={0cm 0cm 0cm 0},clip]{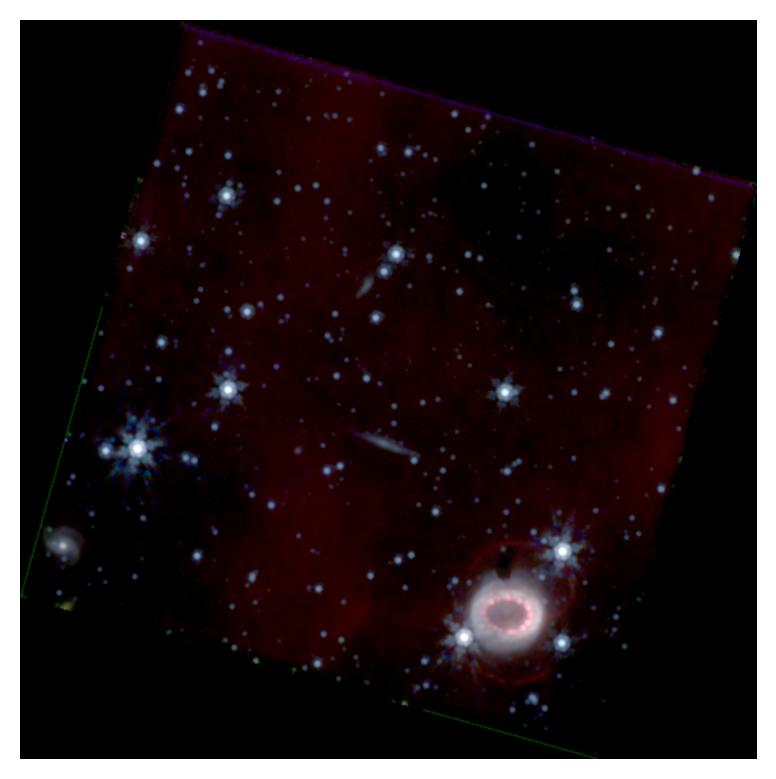}
        \caption{NIRCam long wavelength channel images, covering the entire field of view around SN\,1987A. North is top, east is to the left, and the area covers 25.4'' $\times$ 25.2''.
        The colour allocations are -- blue: F323N, green: F356W, red: F405N, and brown : F444W. 
        \label{fig:whole_images}}
\end{figure}

We chose to use the subarray of NIRCam detectors rather than using the full array, because this observing mode saves total observing time and avoids total saturation by nearby field stars.
SN\,1987A measures approximately 8'' across its outer rings, and the whole system was completely covered by the 320$\times$320 pixel sub-array of the detector by the SUB320 mode of the Module B camera. 
In the SW channel, the total area covered is approximately $25''\times 25''$,
with a $\sim 5''$ gap between the subarrays of the 4 detectors.
In the LW channel, the field of view is 20.45''$\times$20.50'' using
a single subarray (with larger pixels).
Fig.~\ref{fig:whole_images} demonstrates the coverage of SUB320 images of the LW channel.
The NIRCam pixel scales are 0.0317'' in the SW channel and  0.0647'' in the LW channel.
The full width at half maxima (FWHM) of the point spread functions (PSF)  are 0.066'' at F200W and 0.16'' at F356W\footnote{\url{https://jwst-docs.stsci.edu/jwst-near-infrared-camera/nircam-performance/nircam-point-spread-functions}}.
The RAPID reading mode was chosen for wide-band filters, while MEDIUM 8 was used for longer exposures with the narrow-band filters.
The total exposure times on source are summarised in Table\,\ref{observing_log}, ranging from 1155 to 37716\,s per filter band.

The NIRCam sensitivities were approximately 30\,per\,cent better than the pre-flight prediction \citep{Rigby.2022}. As a consequence, we achieved
0.004--0.11 MJy\,sr$^{-1}$ at 1 $\sigma$ noise levels, measured in a blank sky area (Table\,\ref{observing_log}), which constitutes very deep imaging of SN\, 1987A.
Fig.~\ref{fig:whole_images} shows that this deep image can detect Br$\alpha$ emission in the ambient interstellar medium, evident as the patchy background emission in F405N filter (red).

\subsection{Data reduction}

The NIRCam data were reduced using the JWST Calibration Pipeline \citep{2023zndo...7829329B} software version 2022$_-$3. The calibration software version number was 1.8.5 with CRDS version 11.16.16 and CRDS context jwst$_-$1023.pma.
$1/f$ noise was removed. 

Field stars were used to align NIRCam images with different jitter positions, as well as for absolute astrometry.
First, an {\it HST} ACS F656W image observed in 2003 \citep{Tziamtzis.2010} was aligned with 23 \textit{Gaia} stars \citep{Collaboration.20187gh} in the field of view, and the astrometry of the {\it HST} image was corrected.
The {\it HST} F656W image was fed into NIRCam Aperture Photometry python code\footnote{\url{https://spacetelescope.github.io/jdat_notebooks/notebooks/aperture_photometry/NIRCam_Aperture_Photometry_Example.html}} 
which is part of {\sc Jdat$_-$notebooks} to create a point source catalogue within the HST field, with coordinates measured from the HST image.
This catalogue was cross-matched against point sources in all NIRCam images; about 50--100 point sources were detected in common between NIRCam images and the {\it HST} image.
The coordinates of these sources were used to correct image distortion, jittered image frames, and to perform astrometry.
A colour composite of the NIRCam LW images covering the entire field of view around SN\,1987A is shown in Fig.~\ref{fig:whole_images}.

\subsubsection{Image convolutions}

In order to derive colours from two images, the angular resolutions must be matched. For this process, the images were convolved with PSFs.
Simulated PSFs were calculated using WebbPSF \citep{Perrin2014}.
The convolution was processed, using the {\sc python} code {\sc PSF Matching}  \citep{Gordon2008, Aniano2011}.
This module creates a common resolution kernel from the simulated PSFs for each convolution operation between higher- and lower-resolution filters. Window functions in signal processing are generally used to taper out the high-frequency noise generated during Fourier transformations within convolution algorithms \citep{1455106}. Thus, we implemented a Cosine Bell (also called a Hanning) window for convolutions between short and long-wavelength channel filters.  On the other hand, a split Cosine Bell window was used for convolutions among filters within the short and long wavelength channels.

\subsection{Line emission within filter bands}

In order to assess the important line and continuum contributions to filter band passes, we used the spectra from SINFONI \citep{Larsson:2016bj} and  {\it JWST}/NIRSpec \citep{Larsson.2023ytc}.

\begin{figure*}
	 \includegraphics[height=7.1cm, trim={0.75cm 0.2cm 1.6cm 1.2cm},clip]{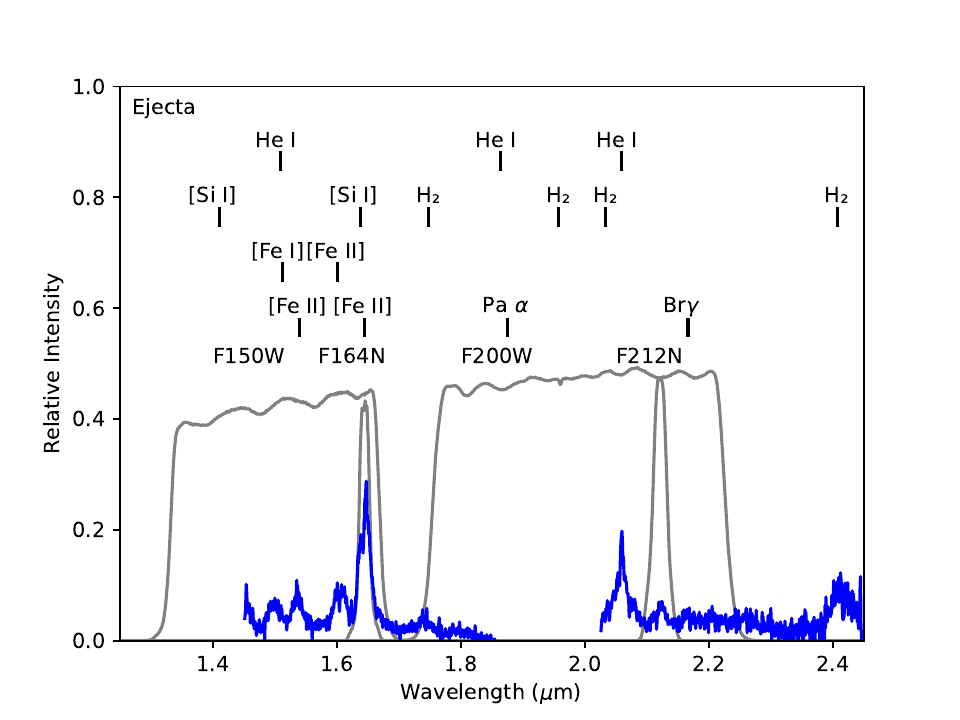}
	 \includegraphics[height=7.1cm, trim={2.0cm 0.2cm 1.6cm 1.2cm},clip]{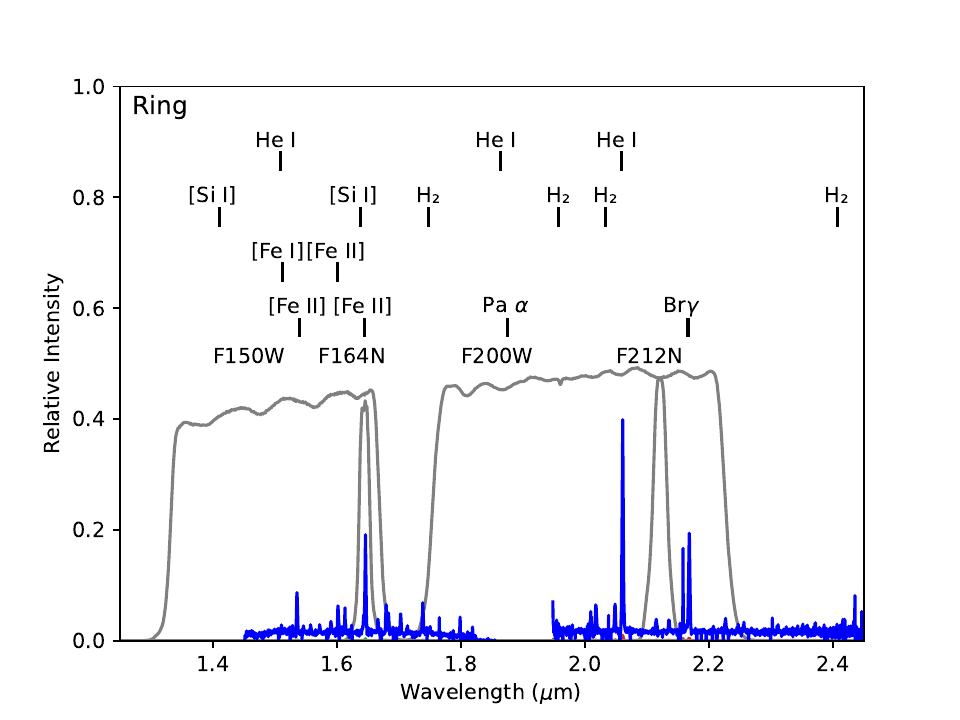}
        \caption{The filter transmission curves of the NIRcam short wavelength channel (grey lines). SINFONI spectra of the ejecta ({\it left}) and ring ({\it right}) in $H$- and $K$-bands \citep{Larsson:2016bj} are also plotted in blue lines, with line-identifications labelled.
        \label{fig:filters}}
\end{figure*}

Fig.~\ref{fig:filters} compares the $H$- and $K$-band spectra of SN\,1987A with the NIRCam filter transmission curves\footnote{\url{https://jwst-docs.stsci.edu/jwst-near-infrared-camera/nircam-instrumentation/nircam-filters}}.
These are pre-flight transmission curves and include instrumental throughput.
The figure also includes VLT/SINFONI \citep{2003SPIE.4841.1548E} spectra of the SN\,1987A equatorial ring and the ejecta in the $H$- and $K$-bands, that were obtained in 2014 October and December 
 \citep[days 10,090--10,152 since the explosion;][]{Larsson:2016bj}.
 The line identifications are taken from \citet{Kjaer.2007} and \citet{Larsson:2016bj,Larsson.2023ytc}.
 H$_2$ line wavelengths are from  \citet{Roueff.2019}.

The broad-band filters, in particular those in the SW channel, contain some strong lines, as detected in the {\it JWST}/NIRSpec observations \citep{Larsson.2023ytc}.  These strong lines are listed in  Table\,\ref{lines}. 
Between the ejecta and the equatorial ring, different lines contribute to filter bandpasses, so that they are listed separately in the Table. 

The F150W band has contributions from  \FeI, \FeII, and \SiI\, lines in the ejecta and \FeII\, lines in the equatorial ring.
The F200W filter has contributions from \Paa, \HeI, and H$_2$ $v$=1--0 S(3) and S(2) lines in the ejecta, while the ring is dominated by atomic lines such as \HI\, (both \Paa\, and \Brg), and \HeI\,lines.

The F164N filter has contributions from both the \FeII\, and \SiI\, lines in the ejecta.
With respect to the 1.64355\,$\mu$m  
(vacuum wavelength) [Fe\,{\small II}] $a^4D_{7/2}$--$a^4F_{9/2}$ line, the half-power bandwidth of the F164N filter covers velocities from $-$1855 to +1426\,\kms, where
the systematic velocity of SN 1987A of +287\,\kms\, \citep{Meaburn.1995} is accounted for.
Although a  majority of the line intensity from the \FeII\, falls within the F164N filter band, the line width extends beyond $\pm$2000\,\,\kms\, \citep{Larsson:2016bj}, so the F164N filter band misses the fast components.
The \SiI\,1.64545\,$\mu$m line also falls in the F164N bandpass, and the filter width corresponds to $-$2203\,\kms\, to 1082\,\kms\, in the rest frame.
In the ejecta as a whole, the SINFONI and NIRSpec spectra show that the line intensity ratio of  \FeII\, to \SiI\, is approximately 1/3 to 1/2 \citep{Jerkstrand:2011fz, Larsson.2023ytc}. 
The ratio of the contribution of the two lines to the intensity in the F164N band might be slightly different at different locations.
In the ring, the 1.644\,$\mu$m \FeII\, line dominates the line intensity, and little contribution from \SiI\,line is present. Similarly, the F212N filter covers the H$_2$ $\nu$=1--0 S(1) line from $-$1432 to 2100\,\kms\, and the F405 filter covers \Bra\, from $-$1319\,\kms\, to 2083\,\kms. The majority of the intensities of these lines are within the filter bands, but the broader (but fainter) components are not covered by these narrow-band filters.

The fluxes of the LW channel have more contribution from the continuum.
The NIRSpec spectra of the ejecta and the equatorial ring show that the continuum rises with wavelength both in the equatorial ring and in the ejecta \citep{Larsson.2023ytc}.

Although the F323N filter was designed to detect the 3.234\,$\mu$m H$_2$ $v$=1--0 O(5) line \citep{Black:1987p92},
the image of SN~1987A in this filter is similar to those made using the broad band filters of the LW channel, in particular F356W.
We conclude that the F323N flux is dominated by the continuum rather than the H$_2$ line in all features, except for faint crescent-shaped structures observed between the ejecta and the equatorial ring. This will be explored further in Matsuura et al. (in preparation).



\begin{table*}
\caption{{\it JWST}/NIRCAM observations of SN\,1987A.  \label{observing_log} }
\begin{filecontents*}{observing_log.csv}
Channel,Filter,Filter,,Pivot,BW,lambda_blue,Lambda_red,PSF empirical FWHM,Exposure time,BK,Sigma (BK),Error,Ejecta lines,Ring lines,Notes
Short,F150W,F150W,,1.501,0.318,1.331,1.668,0.050,1347,0.175,0.067,0.134,[Fe {\small I}] 1.444;  [Si {\small I}] 1.607;  [Fe {\small II}]+[Si {\small I}] 1.64 blend,[Fe {\small II}]  1.534; [Fe {\small II}]  1.600; [Fe {\small II}]  1.644,
,F164N$^*$,F164N,F150W,1.645,0.020,1.635,1.653,0.056,37716,0.802,0.112,0.059,[Fe {\small II}]+[Si {\small I}] 1.64 blend,[Fe {\small II}]  1.644,
,F200W,F200W,,1.989,0.457,1.755,2.226,0.066,1155,0.235,0.081,0.131,H {\small I} 1.875 (Pa$\alpha$);  He {\small I} 2.059; H$_2$ 2.122,H {\small I} 1.875 (Pa$\alpha$); He {\small I} 2.059; H {\small I} 2.166 (Br$\gamma$),
,F212N,F212N,,2.121,0.027,2.109,2.134,0.072,28287,0.782,0.084,0.058,H$_2$ 2.122 $v$=1–0 S(1),He {\small I} 2.114; [Fe {\small II}] 2.135,
Long,F323N$^*$,F323N,F322W2,3.237,0.038,3.217,3.255,0.108,28287,-0.015,0.007,0.050,(H$_2$ $v$=1--0 O(5)),,Continuum
,F356W,F356W,,3.568,0.781,3.140,3.980,0.116,1347,-0.418,0.004,0.067,H$_2$ 3.846; Unidentified $\sim$3.88$\mu$m,H {\small I} 3.297; H {\small I} 3.744,Lines weak relative to continuum
,F405N$^*$,F405N,F444W,4.052,0.045,4.028,4.074,0.136,37716,0.485,0.071,0.034,H {\small I} 4.052 (Br$\alpha$),H {\small I} 4.052 (Br$\alpha$),
,F444W,F444W,,4.408,1.029,3.880,4.986,0.145,1155,-0.020,0.014,0.066,H {\small I} 4.052 (Br$\alpha$); H {\small I} 4.654;  H$_2$  4.695,H {\small I} 4.052 (Br$\alpha$); H {\small I} 4.654,
\end{filecontents*}
\csvreader[tabular= l l c c c c c rr ll lll,
				table head=\hline Channel &  Filter  & $\lambda_p$ & BW & $\lambda_ B$ & $\lambda_R$ & PSF & $t_{\rm exp}$ & $I_{\nu\, {\rm BK}}$ & $\sigma I_{\nu {\rm BK}}$ & Error  \\ \hline\hline, 
				table foot=\hline ] 
				{observing_log.csv}
				{} 
				{\csvcoli & \csvcolii   & \csvcolv  & \csvcolvi & \csvcolvii    & \csvcolviii  & \csvcolix  & \csvcolx & \csvcolxi & \csvcolxii & \csvcolxiii} 
\\
$\lambda_p$: pivot wavelength in $\mu$m \citep{Tokunaga.2005}.
BW: bandwidth in $\mu$m. 
$\lambda_ B$ and $\lambda_R$: the half power wavelengths in $\mu$m of a passband at blue and red wavelengths, at which the transmission falls to 50\,per\,cent of its peak value. 
\footnote{\url{https://jwst-docs.stsci.edu/jwst-near-infrared-camera/nircam-instrumentation/nircam-filters}}
PSF: Full-width half maximum (FWHM) of the empirical point spread function.
\footnote{\url{https://jwst-docs.stsci.edu/jwst-near-infrared-camera/nircam-performance/nircam-point-spread-functions}}
$t_{\rm exp}$: exposure time on source in sec.
$I_{\nu, {\rm BK}}$ and $\sigma I_{\nu {\rm BK}}$: mean and 1 $\sigma$ of `blank sky' level in a nearby field (MJy\,sterad$^{-1}$).
Error: mean error level estimated from the error map in a nearby field but off from SN~1987A. This error combines Poisson noise and read noise (MJy\,sterad$^{-1}$).
$^*$F164N, and F323N and F405N filters are used together with the wide band filters F150W, F322W2 and F444W, respectively.
\end{table*}


\begin{table*}
\caption{The strong lines falling into the NIRCam filter bands (with their vacuum wavelength in $\mu$m).  Lines are listed separately for the ejecta and the equatorial ring, and are taken from  \citet{Larsson.2023ytc}. \label{lines} }
\csvreader[tabular= l l  lll,
				table head=\hline Channel &  Filter  & Ejecta  & Ring  & Note \\ \hline\hline, 
				table foot=\hline ] 
				{observing_log.csv}
				{} 
				{\csvcoli & \csvcoliii   & \csvcolxiv & \csvcolxv & \csvcolxvi } 
\\
\end{table*}


\subsubsection{Auxiliary data: ALMA}

ALMA observed SN\,1987A with two tunings, 301.75--305.57\,GHz in Kinematic Local Standard of Rest (LSRK)  (993.51--981.09\,$\mu$m) and 313.73--317.52\,GHz  (955.57--944.17\,$\mu$m), on 2021 November 2nd as part of the Project 2021.1.00707.S.
The total integration time was 10,813\,sec.
The synthesized beam was 0.081"$\times$0.061" with a position angle of 33.7$^{\circ}$ for the major axis.
The shortest baselines allowed for a maximum recoverable scale of 0.993". This implies that the diffuse emission, which extends more than 1", is likely to be over-resolved, resulting in some flux loss.

\section{Results and discussions}

SN\,1987A contains three major components which we describe separately, namely,  the inner ejecta (Sect.\,\ref{sect-ejecta}), the equatorial ring and associated diffuse emission (Sect.\,\ref{sect-outside} and \ref{Sect:power-law}), and the outer rings (Sect.\,\ref{sec-outer-ring}).
Fig.~\ref{fig:three-color} highlights these features using the NIRCam images.

The main findings of the NIRCam images are (1) the substructure of the ejecta, namely, the bar (Sect.\,\ref{sec-bar}), (2) crescents, faint emission between the ejecta and the equatorial ring, (3) the first identifications of the NIR synchrotron emission from the hot spots, which are bright spots composing of the equatorial ring, the outer spots and diffuse emission, which are exterior to the equatorial ring (Sect.\,\ref{sect-outside} and \ref{Sect:power-law}).
The analysis of crescents will be reported in a separate paper.

\subsection{Ejecta} \label{sect-ejecta}

\begin{figure*}
	 \includegraphics[width=8.8cm, trim={1.1cm 1.1cm 0cm 1.1cm},clip]{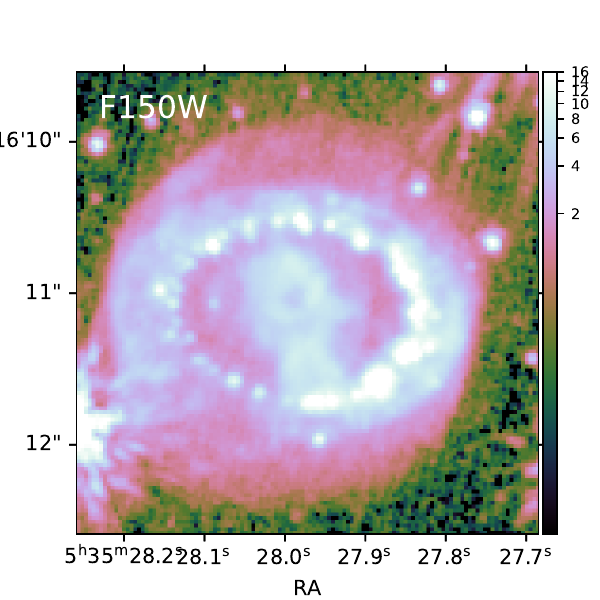}
	 \includegraphics[width=8.8cm, trim={1.1cm 1.1cm 0cm 1.1cm},clip]{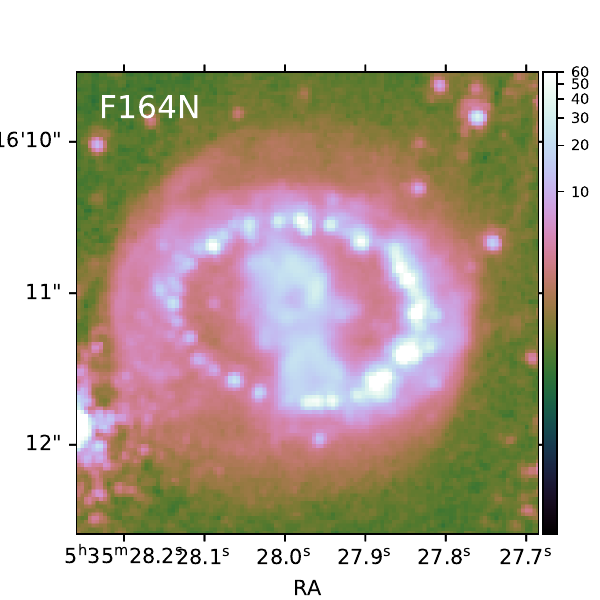}
	 \includegraphics[width=8.8cm, trim={1.1cm 1.1cm 0cm 1.1cm},clip]{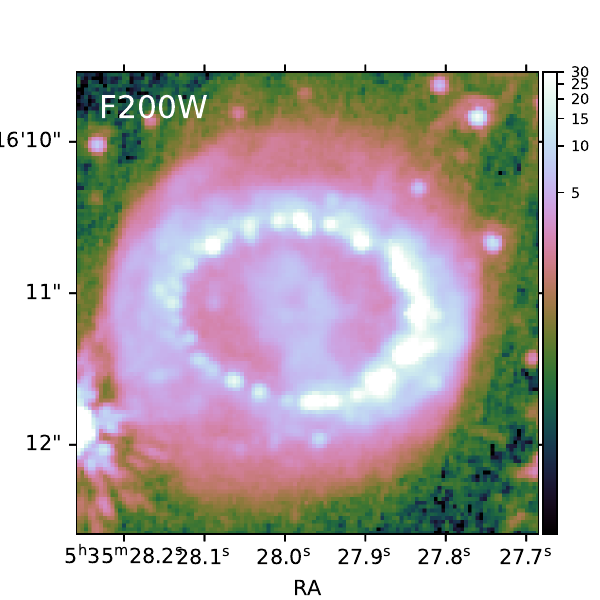}
	 \includegraphics[width=8.8cm, trim={1.1cm 1.1cm 0cm 1.1cm},clip]{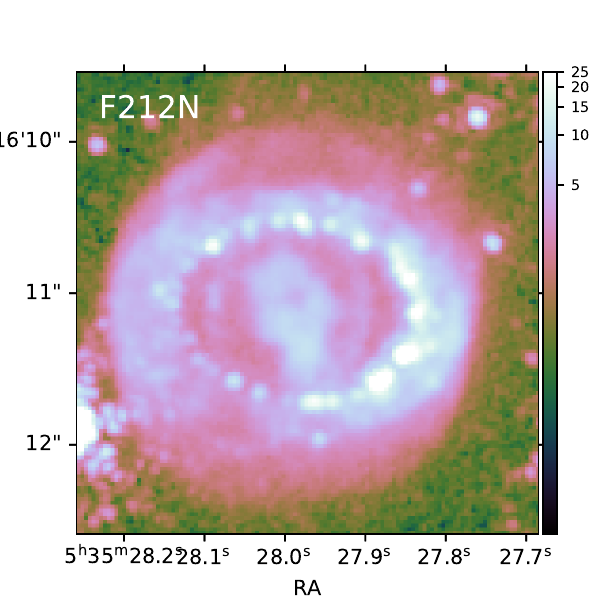}
        \caption{Short wavelength channel images of the equatorial ring and the faint extended emission beyond the ring and ejecta.
        Within the ejecta, a bar, extending from the east to the west is found. The equatorial ring consists of more than 20 hot spots, while over 10 outer spots exterior to the equatorial ring are detected. Faint diffuse emission extends far beyond the outer hot spots. The coverage of the images is 3''$\times$3'', with the north at the top, and the colour stretch is in the unit of MJy~sr$^{-1}$. 
        \label{fig:short_images}}
\end{figure*}

The ejecta are detected in all four filter band images in the SW channel (Fig.\,\ref{fig:short_images}).
The keyhole or hourglass shape of the ejecta extends from the north to the south. 
The ejecta contain a dip or  `hole' near the centre with a slight offset to the north.
This `hole' is not a hole in the gas, but probably due to the heavy extinction (self-absorption) by recently formed SN ejecta dust \citep{Matsuura:2011ij, Indebetouw:2014bt, Cigan:2019cl}.
At the waist of the ejecta,  a small bar is seen from the east to the west in blue colour in Fig.\ref{fig:three-color} and F150W and F164N images (Fig.\ref{fig:short_images}). 
The southern part of the ejecta breaks into two branches near the equatorial ring.

\begin{figure*}
	 \includegraphics[width=8.8cm, trim={1.1cm 1.1cm 0cm 1.1cm},clip]{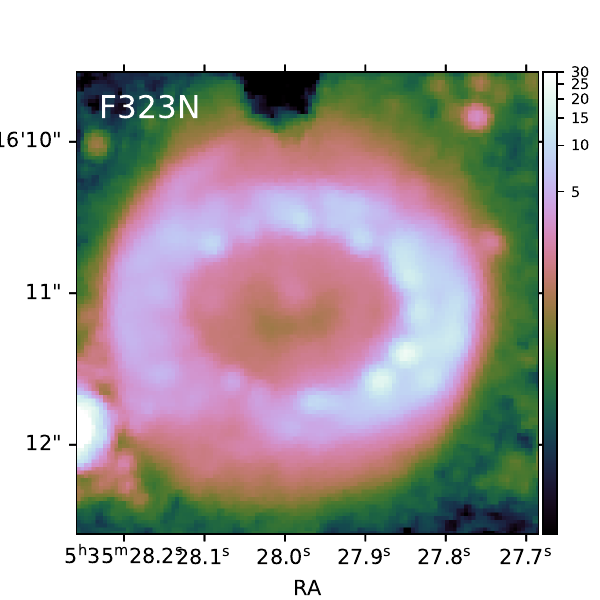}
	 \includegraphics[width=8.8cm, trim={1.1cm 1.1cm 0cm 1.1cm},clip]{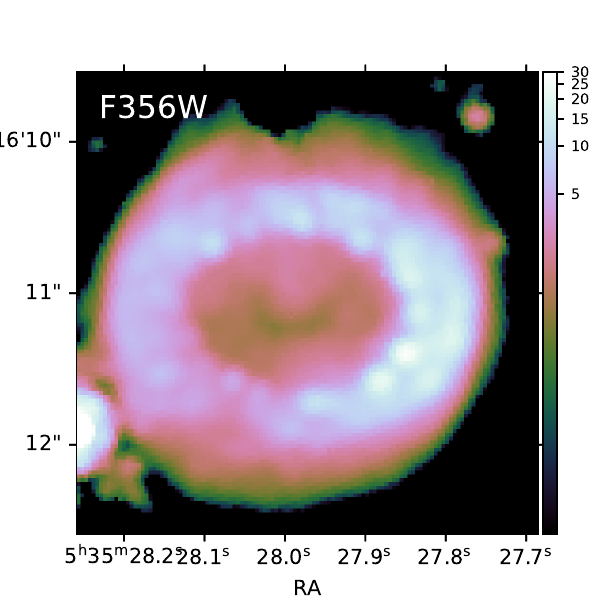}
	 \includegraphics[width=8.8cm, trim={1.1cm 1.1cm 0cm 1.1cm},clip]{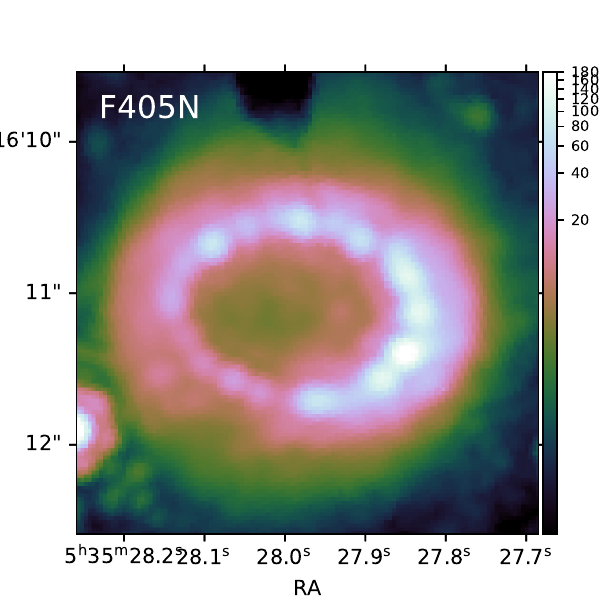}
	 \includegraphics[width=8.8cm, trim={1.1cm 1.1cm 0cm 1.1cm},clip]{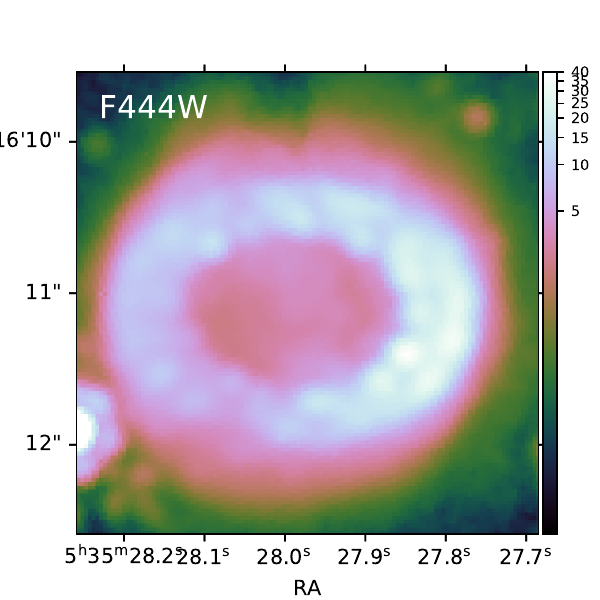}
        \caption{ NIRCam LW channel images. The outer spots exterior to the equatorial ring are as bright as the hot spots within the ring in the LW images, apart from F405N. 
The ejecta emission in the LW images is dominated by continuum, mostly due to dust. The dip in the centre of F343N, F356W and F405N is caused by dust self-absorption. Only two blobs are detected in these images. The F405N image traces both the dust continuum and Br$\alpha$ in the ejecta.
The coverage of the images is 3''$\times$3'', with the north at the top, and the colour stretch is in the unit of MJy~sr$^{-1}$.
        \label{fig:long_images}}
\end{figure*}

On the other hand, the LW images do not show a clear keyhole shape in the ejecta (Fig.\,\ref{fig:long_images}).
Instead, there are two faint blobs in the north and the south in the F323N and F356W images.
In fact, the northern blob in the F323N and F356W images corresponds to the hole in the SW channel images.
In the middle of the two blobs, a ``gap'' spreads from the east to the west.
The exception is the F444W image, which shows a hint of the keyhole shape and the bar, similar to that seen in the SW images.

\begin{figure*}
	 \includegraphics[height=8.cm, trim={4.5cm 1.5cm 2.5cm 1.7cm},clip]{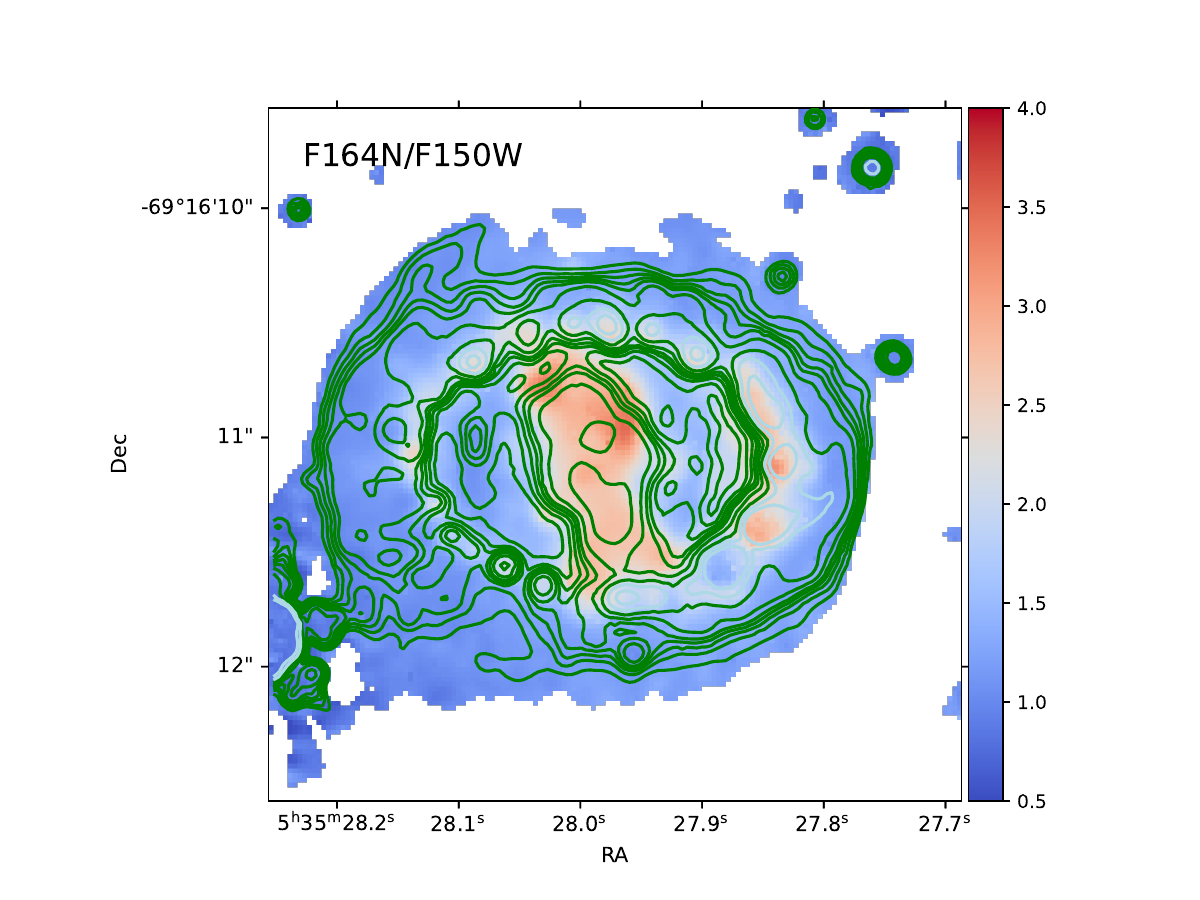}
	  \includegraphics[height=8.cm, trim={4.5cm 1.5cm 2.5cm 1.7cm},clip]{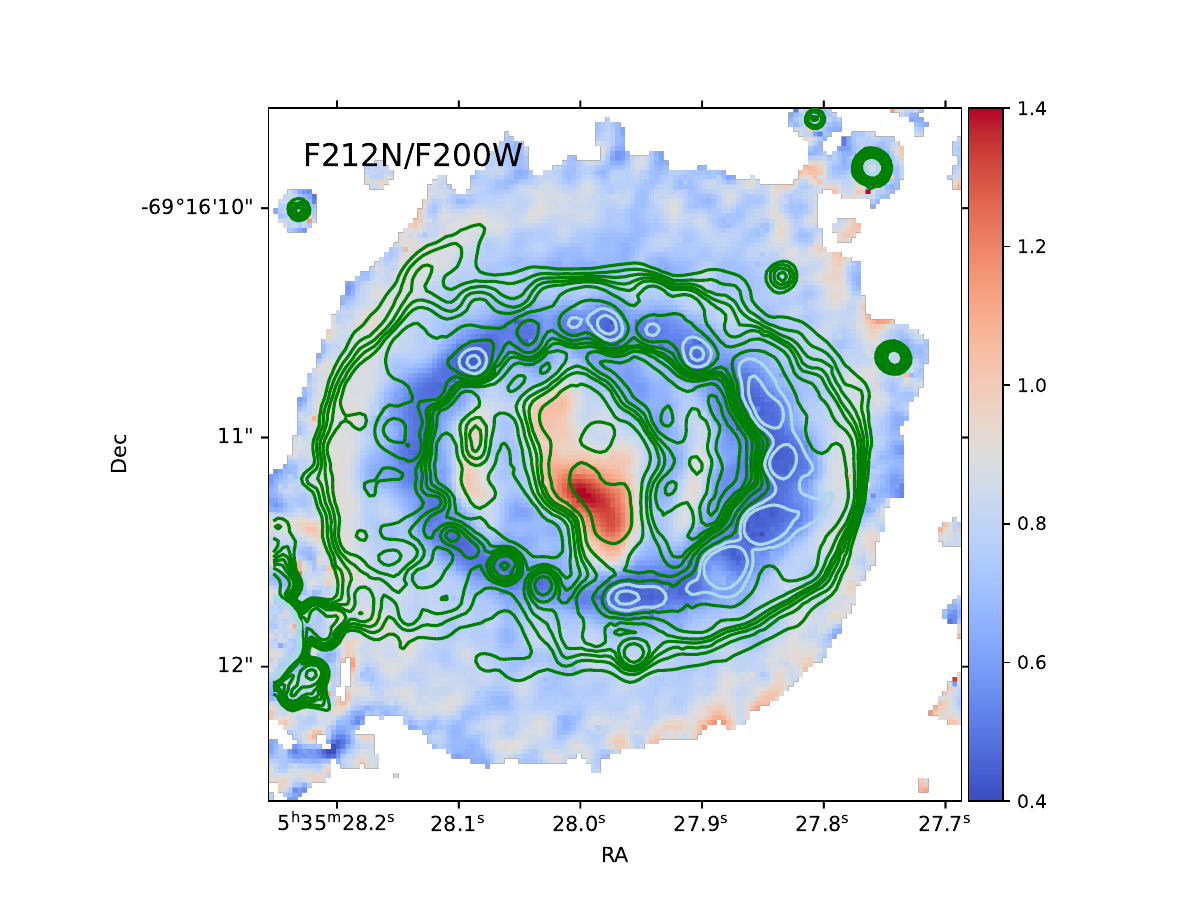}
	  \includegraphics[height=8.cm, trim={4.5cm 1.5cm 2.5cm 1.7cm},clip]{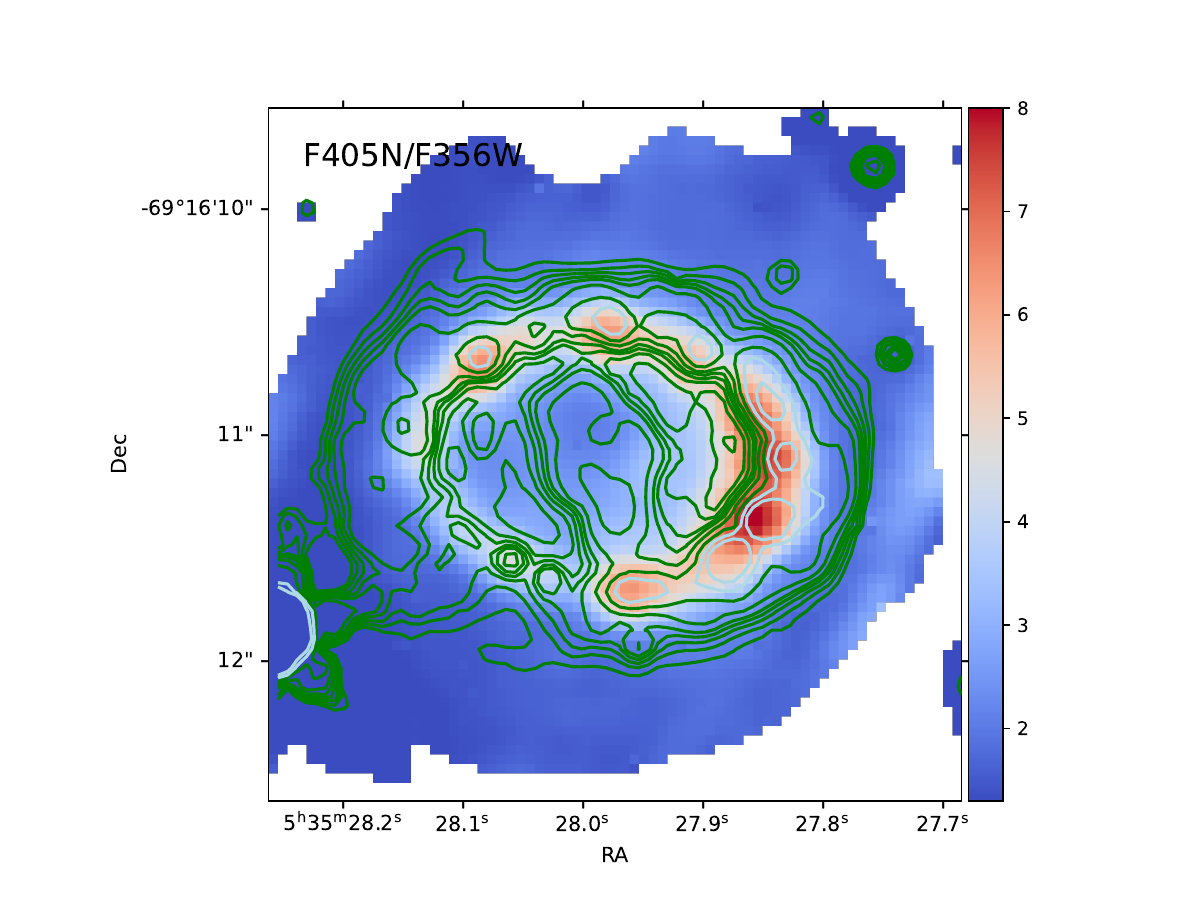} 
	  \includegraphics[height=8.cm, trim={4.5cm 1.5cm 2.5cm 1.7cm},clip]{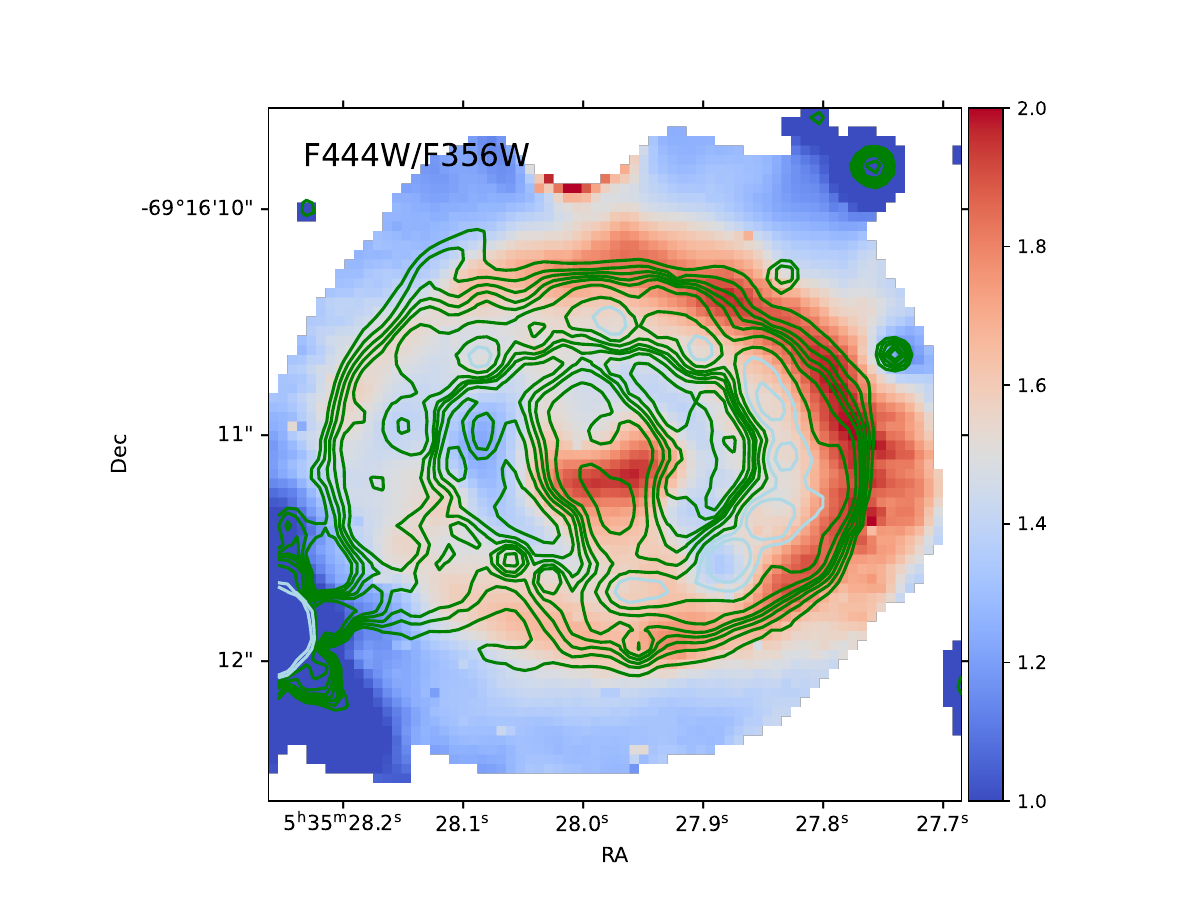} 
        \caption{The colours show the intensity ratios of two filter bands. The line contours are plotted to guide the eye; these represent the F212N image, smoothed to the angular resolutions of the filter bands used to compute the intensity ratios. 
        \label{fig:color}}
\end{figure*}

The ratio of F444W/F356W in Fig.\,\ref{fig:color} (right bottom) indicates an excess in F444W in the bar. 
This suggests that the dust self-absorption towards the bar is still strong at F356W, while the dust self-absorption is less in F444W.

While the SW bands have emission line contributions, 
the ejecta emission in the LW bands is dominated by continuum emission \citep[see Sect.~\ref{Sect:power-law} and Fig.~\ref{fig:SED}; also][]{Larsson.2023ytc}.
The NIRCam SED decomposition analysis shows that in the LW bands the ejecta fluxes are indeed dominated by the continuum \citep{2023arXiv230913011A}.
The source of the continuum is presumably from cold SN ejecta dust \citep{Matsuura:2011ij, Indebetouw:2014bt, Matsuura:2015kn}. The ALMA continuum images at 450\,$\mu$m and 870\,$\mu$m show more dust towards the centre of the system \citep{Cigan:2019cl}. This causes dust self-absorption around the waist of the ejecta at  NIR wavelengths, which is seen as a dip at the centre of the ejecta in NIRCam images.  
As dust extinction is higher at the shorter wavelength of the NIR, the SW images of this area trace mainly the foreground of the heavily-obscured dusty regions in the ejecta.

The dust emission extends across the ejecta \citep{Cigan:2019cl},
but is fainter towards the north and south edges of the ejecta,  causing less extinction at NIR wavelengths.
These regions are also heated by the UV radiation from the equatorial ring, hence, the NIR emission is detected as blobs in the LW images.
The total flux of the ejecta is about 12\,$\mu$Jy at F356W. Extrapolating the cold ($\sim$20\,K) dust emission from the millimetre wavelengths to the NIR would result in less than 1\,$\mu$Jy at F356W. Ejecta dust must be heated to temperatures higher than 20\,K in the blobs to emit 12\,$\mu$Jy in this band.
The obscuration of the hole is seen because cold dust in the centre absorbs the emission from the dust in the far end of the warmer ejecta.


\subsubsection{Bar}\label{sec-bar}

The bar is a small structure,  sticking out to the east and to the west at the waist of the ejecta (Figs.~\ref{fig:three-color} and \ref{fig:short_images}). 
The east component of the bar is slightly tilted to the southeast.
The west side of the bar has been detected by {\it HST} H$\alpha$ imaging before \citep[e.g.][]{Larsson:2013gx}, but
in the optical images,  the east side of the bar was not detected clearly. 
This indicates dust extinction in the line of sight on the east side of the bar.

The apparent structure of the bar slightly differs from one wavelength to another (Fig.~\ref{fig:short_images}). In the F164N image, the bar looks like a triangular shape, 
separated from the main body of the ejecta.
The bar is also found in the F150W and F200W images, but less distinct than in F164N.
In the F212N  image, there is only a faint trace of the bar structure. 
This shows that the emission causing the bar is mainly from the atomic lines such as \FeII\ + \SiI\ (F150W and F164N) and probably also \FeI\ at 1.44\,$\mu$m in F150W, or \HI\ and \HeI\ (F200W), but not H$_2$ (F212N).

\begin{figure}
	 \includegraphics[width=0.47\textwidth, trim={1.1cm 1.0cm 0cm 1.cm},clip]{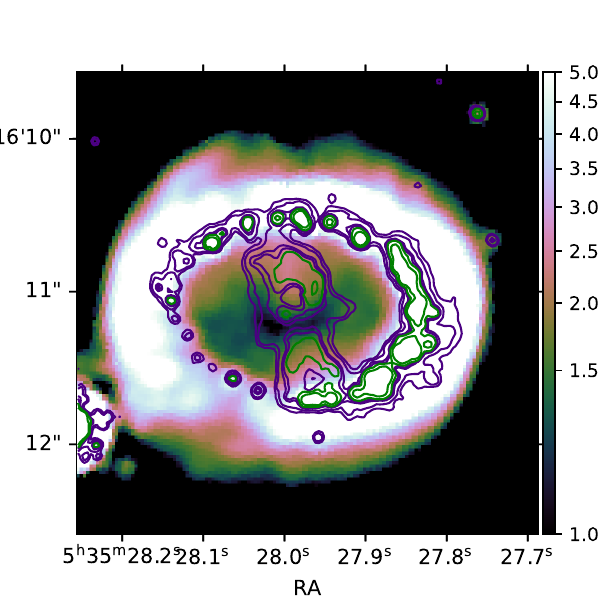}
        \caption{ F356W image in colour contours in the unit of MJy~sr$^{-1}$, with F164N represented by line contours. %
        It looks like that the bar emission in F164N is a dip in intensity in the F356W image.
          \label{fig:F356W_F164N}}
\end{figure}

\begin{figure*}
	 \includegraphics[width=8.8cm, trim={1.2cm 1cm 1cm 1.1cm},clip]{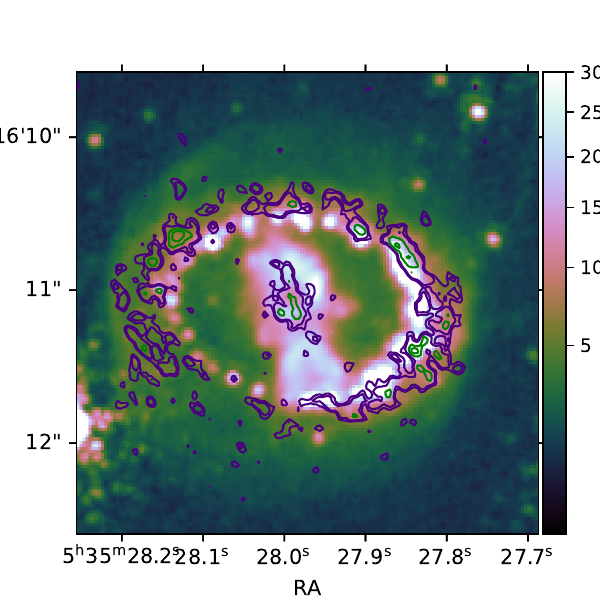}
	 \includegraphics[width=8.8cm, trim={1.2cm 1cm 1cm 1.1cm},clip]{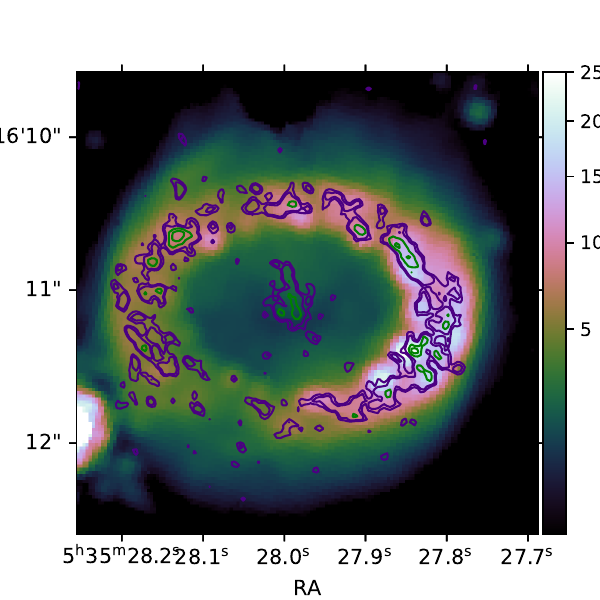}
        \caption{F164N image ({\it left}) and F356W image ({\it right}) with ALMA 315\,GHz continuum emission shown as line contours. 
        The ALMA  continuum image contains synchrotron radiation in the ring and dust thermal emission in the ejecta. The hot spots and the outer spots detected in the F356W image generally have corresponding spots in ALMA 315 GHz synchrotron emission.
          \label{fig:comparison_images_ALMA}}
\end{figure*}

At the location of the bar, there is a dip in surface brightness in F323N and F356W (Fig.\,\ref{fig:long_images}). 
Fig.\,\ref{fig:F356W_F164N} shows the F356W image in colour, with contours of the F164N image.
The bar in the F164N image corresponds to the location of the low surface brightness region in the F356W image.
The waist of this bar also contains some emission of the dust in the ALMA 315\,GHz image (Fig.~\ref{fig:comparison_images_ALMA}).
This region probably contains a large mass of dust, so that it absorbs and scatters near-IR light. 
If there is substantial IR extinction, the \FeII, \SiI, \HI, and \HeI\, emission must come from the surface of the bar structure on the near side to us, and therefore does not reflect the whole structure of the bar.


\subsubsection{\SiI\, and \FeII\,emission}

In general, the 1.64\,$\mu$m \FeII\, line is often used as a tracer of shocks. In SN remnants there is the complication of blending of two lines (\SiI\, and \FeII) in the F164N filter.
These two atomic stages have quite different ionisation potentials.
Fe$^+$ has an ionisation potential of  16.2\,eV \citep{Mouri.2000}.
The ionisation energy of Si$^0$ is only 8.2\,eV. 

The top left panel of Fig.\,\ref{fig:color} shows 
the ratio of the F164N image to the F150W image (\Fii/\Fi). 
The interpretation of this image requires some caution, because the F150W image, which is typically expected to be a continuum, actually contains atomic lines, such as \FeI. 
Nevertheless, the ratio image combined with the F164N image should show characteristics of the  \SiI\, and \FeII\, emitting regions,   as the F164N intensity is typically three times brighter than the F150W intensity in the ejecta.

The \Fii/\Fi\, ratio in Fig.~\ref{fig:color} shows two peaks, one at the north-northeast and the other at the northwest edges of the ejecta.
The primary peak at the northwest ejecta is found in the original F164N image, so that this is due to an excess of \SiI\, and \FeII\, line intensities, and not due to the dip in the \FeI\, line intensity in F150W.


While the 1.64\,$\mu$m \FeII\, line is often used as a tracer of shocks in SNRs or AGNs, 
 this line in SN\,1987A might not be solely due to shock excitation.
The excitation mechanisms of lines might be different in the outer and the inner parts of the ``inner ejecta''. (Note that historically, ``outer ejecta'' refers to the ejecta with high velocity but low density, extending beyond the keyhole, and some part of this has passed the equatorial ring and has been interacting with the ring and its exterior material.) 
The prominent \FeII\, line at 1.644\,$\mu$m is due to the $a^4F_{9/2}$--$a^4D_{7/2}$ transition \citep{Mouri.2000}.
The critical density for collisional deexcitation is $n_{\rm crit}\sim10^5$\,cm$^{-3}$ at the electron temperature of $T_{\rm e}$=$10^4$\,K 
\citep{Shull.19821zp, Greenhouse.1991}.
At lower kinetic temperature and lower density, the excitation of the \FeII\, 1.644\,$\mu$m line is mainly by UV fluorescence and non-thermal excitation and ionization, followed by radiative de-excitation to the upper $a^4F_{9/2}$ level of the 1.644\,$\mu$m line \citep{Jerkstrand:2011fz}. Thermal collisional excitation is not important in the inner region: it has a very low temperature, only $< 200$\,K even at 8 years after the explosion \citep{Jerkstrand:2011fz}, compared to the excitation temperature of the $a^4F_{9/2}$ level of 11,445\,K.

However, thermal collisional excitation may be important for the \FeII\,  line in the region close to the ring, where the X-ray photoionization and heating \citep{2013ApJ...768...88F} may result in a temperature $> 2000$\,K, enough to excite this line.
The secondary peak of the emission at the north-northeast of the ejecta close to the equatorial ring (the top left panel of Fig.\,\ref{fig:color}) probably traces the interaction of the ejecta with the equatorial ring at that location \citep{Larsson.2023ytc}.

Integrated across the entire ejecta, the \SiI\, line is predicted to contribute more to the F164N intensity than the \FeII\, line \citep{Larsson.2023ytc}.
 At the primary peak, instead of \FeII,  the \SiI\, line might contribute more to the intensity, and its excitation could be due to recombination at temperatures below 200\,K  
\citep{Raymond.2018, Larsson.2023ytc}.

\subsubsection{H$_2$ emission}

The F212N image in Fig.\,\ref{fig:short_images} and the \Fiv/\Fiii\, ratio in Fig.\,\ref{fig:color} show the H$_2$ emitting region in the ejecta. 
The overall shape of the H$_2$ emitting region within the ejecta is similar to those of the \Ha\,  \citep[F625W; ][]{Larsson.2019h9} and \FeII\, emitting regions (Fig.\,\ref{fig:short_images}), but the H$_2$ peak is located in the south, corresponding to the locations where the \Ha\, and \FeII\, emissions are fainter.
The H$_2$ strength has almost an anti-correlation with the \FeII\, strength: the peak of \FeII\, is in the north-west, and that side of the ejecta has weaker H$_2$ emission.
The anti-correlation can also be found in \Ha\, and H$_2$ \citep{Larsson.2019h9}.
This could be the effect of the strength of the X-ray and 912--1100\,\AA\, UV illumination from the equatorial ring onto particular locations within the ejecta. 
On the west side of the equatorial ring and just outside the ring, there are on-going strong shocks, while on the south-east side, only weak emission is found from the equatorial ring \citep[Sect.\,\ref{sect-outside}; also][]{Frank:2016ka}. 
The X-ray radiation from the southeast of the ring imprinting onto the southeast part of the ejecta is also weaker, hence, there is more molecular gas found in the southeast than in the rest of the ejecta.

\subsubsection{Br$\alpha$ emission} \label{sect-bralpha}

The bottom left image of Fig.\,\ref{fig:long_images} is the Br$\alpha$-dominated F405N image, and the \Fvii /\Fvi\, ratio in Fig.\,\ref{fig:color} shows the \Bra\, emitting region in the ejecta.
The distinct key-hole shape of the ejecta, found in Fig.\,\ref{fig:short_images} and the {\it HST} H$\alpha$ image \citep{Larsson.2019h9}, is not obvious in \Bra. This is an illusion of the colour stretch used in Fig.\,\ref{fig:long_images}, and there is a slight dip in the brightness.
As seen in Fig.\,\ref{fig:long_images}, Br$\alpha$ is very faint in the ejecta, compared with the equatorial ring. 
Nevertheless, the peak of the Br$\alpha$ emission on the west side of the ejecta is part of the bar.
Very faint \Bra\, emission from the north of the ejecta is also identified as a blob.
It is not clear whether the additional very faint blob on the east side of the ejecta is also due to the bar, with a possible additional contribution from the outer ring.

In the LW channel images, the F405N image is distinctly different from the rest of the images (Fig.\,\,\ref{fig:long_images}). In the F323N and F356W images, there are two separate blobs, one in the north and the other in the south of the ejecta.
In the F405N image, the north blob is detected, but the south blob is missing.
It seems that the Br$\alpha$ emission from just inside the equatorial ring is stronger than the Br$\alpha$ emission from the south blob, burying the emission from the south blob.

\subsection{The equatorial ring and its exterior} \label{sect-outside}

The NIRCam images of SN\,1987A captured detailed structures found within 
and exterior to 
the equatorial ring (Fig.\ref{fig:three-color}).
The equatorial ring is composed of at least twenty-two hot spots along its circumference.
Hot spots are bright parts within the equatorial ring in UV and optical, and in the past, associated with the forward shocks colliding into dense clumps \citep{Sonneborn:1998p29919}. 
They have the highest density within the circumstellar material.
These hot spots within the equatorial ring are the brightest structures within the SN\,1987A system in the SW images (Fig.\,\ref{fig:short_images}).

\begin{figure}
	 \includegraphics[width=0.48\textwidth, trim={0.1cm 0.3cm 0cm 0.0cm},clip]{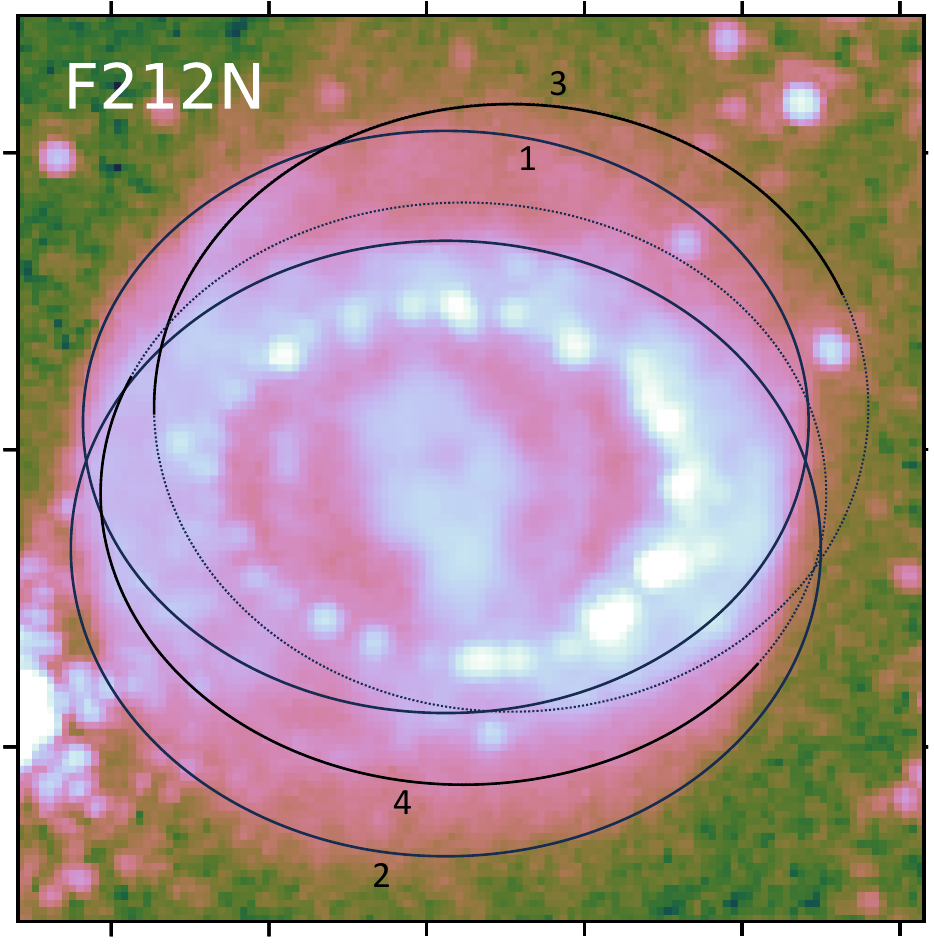}
	         \caption{Illustrating the three or four faint ellipses within the diffuse emission. The ellipses 1 and 2 have well-defined edges, while ellipses 3 and 4 have partial ellipses, and their full structures are not clear. Ellipse 4, tracing on the south part of the diffuse emission, might be tracing a few outer spots, rather than a full ellipse, so that the total number of ellipses could be three  or four.  
        \label{fig:diffuse-rings}}
\end{figure}

Exterior to the equatorial ring, there are rings of faint diffuse emission. 
The outer diffuse emission is composed of at least three ellipses, possibly four, though these numbers and shapes are subjective (Fig.~\ref{fig:diffuse-rings}).
Three separate arcs are clearly seen in the northeast part within the diffuse emission (Fig.\,\ref{fig:short_images}). 
It is unclear whether ellipses 2 and 4 are actually a single ellipse or two separate ellipses that are offset in the north and south directions.
The diffuse emission is expected to be due to the reverse shock \citep{Larsson.2023ytc}. {\it JWST}/NIRSpec observations of the \ion{He}{I} 1.083\,$\mu$m line were used to determine the 3D morphology of the line, which revealed a bipolar shape with bubble-like structures extending outwards on both sides of the equatorial ring \citep{Larsson.2023ytc, 2020A&A...636A..22O}. Projection and limb brightening of the shock surface may create several elliptical structures in the images.

While the faster ($>$ 10,000 km\,s$^{-1}$) but low-density outer ejecta expands continuously since the SN explosion \citep{Larsson:2016bj, Kangas.2021}, it interacts with the circumstellar envelope. 
Until about 2014, the locations of the main interactions of the outer SN ejecta with the circumstellar material were within the equatorial ring \citep{France.2015}. This interaction is now most actively occurring at the equatorial ring on the north and west sides and the exterior to the equatorial ring. The interaction causes emission, which contains continuum emission (a combination of synchrotron and dust, discussed in the next section) and line emission, and appears as diffuse emission outside of the equatorial ring. Diffuse emission is detected by {\it HST} \citep{Larsson.2019h9} but it is clearer in {\it JWST} NIRCam images, because of the brighter contribution of synchrotron emission in the NIR than in the \Ha\ emission observed with {\it HST}.

In these diffuse rings, we found spots, which we refer to as ``outer spots'' to distinguish them from the hot spots within the equatorial ring \citep{Larsson.2019h9}. 
These outer spots and hot spots are spatially unresolved at the NIRCam angular resolutions (FWHM of $<$0.05"; Table\,\ref{observing_log}).
 Having outer spots within the diffuse ring suggests that the mass-loss material from the red supergiant may not have been smooth, but had contained many clumps within. As the thin but faster SN ejecta expands and collides with the circumstellar material, the gas in these clumps is now ionised and shocked, and forms as ``outer spots''.

Fig.\,\ref{fig:long_images} shows that the outer spots are as bright as the hot spots, or even brighter on the east side in the LW images, except for in the F405N image. 
These images are in contrast to the SW images, which show brighter hot spots than the outer spots. The outer spots in the LW images appear to form an outer portion of the equatorial ring, extending all around it \citep{2023arXiv230913011A}. 
In general, both the hot spots and the outer spots are brighter on the western side than on the eastern side.

\subsection{Power-law spectra at 3--4\,\texorpdfstring{$\mu$}{µ}m in the equatorial ring and its exterior}\label{Sect:power-law}

\begin{figure}
		 \includegraphics[width=0.48\textwidth, trim={3.5cm 1.2cm 3cm 1.3cm},clip]{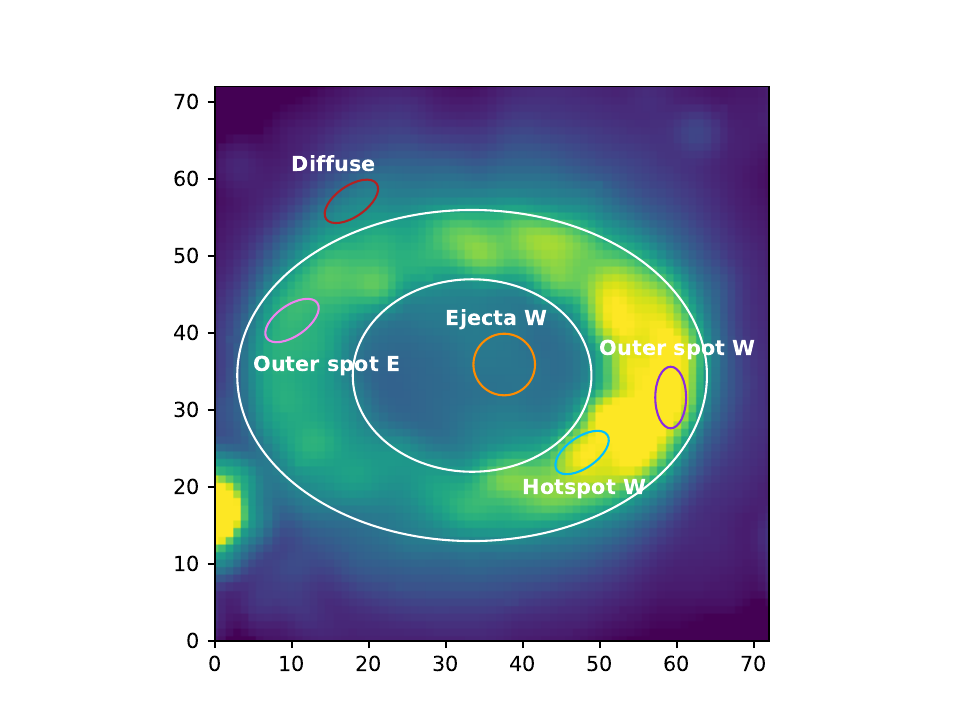}
        \caption{Small regions (in colour) used for SEDs in Fig.\,\ref{fig:SED}  and the white elliptical annuals used for the ring and outer spots photometry in Fig.\,\ref{fig:radio} (also in Sect.\,\ref{app-flux-measurements}), overlaid on F323N image  
                \label{fig:SED-region}}
\end{figure}

\begin{figure}
		 \includegraphics[width=0.5\textwidth, trim={0.4cm 0cm 1cm 1.3cm},clip]{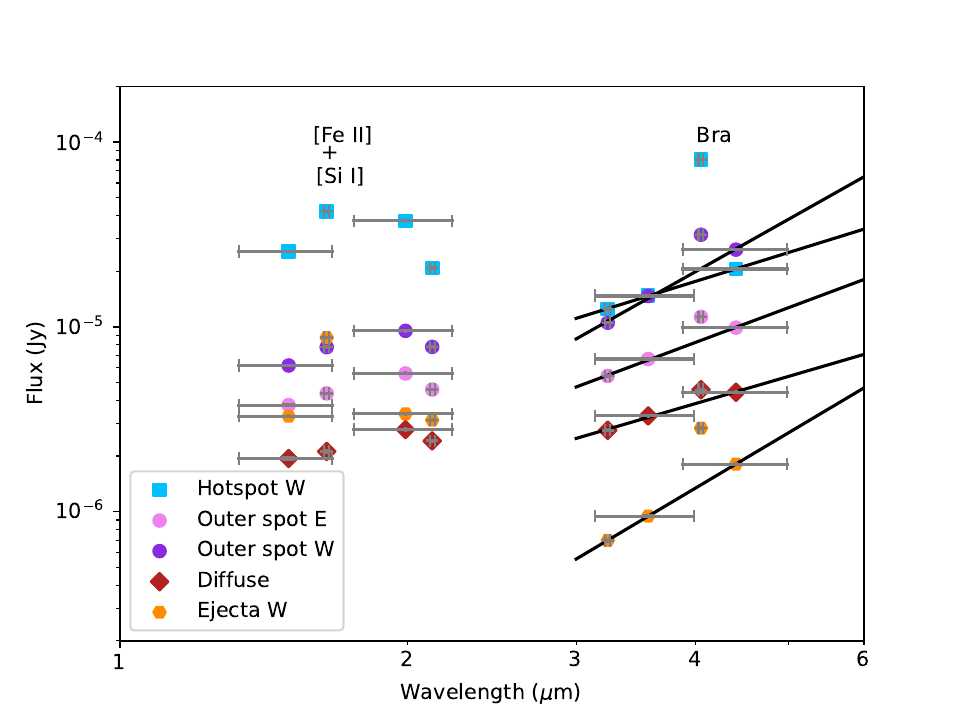}
        \caption{The spectra at represented regions indicated in Fig.~\ref{fig:SED-region}. The black lines are power-law fits to the F323N, F356W and F444W fluxes and the grey lines show the filter widths. The uncertainties of the fluxes are smaller than the plotting symbols. 
                \label{fig:SED}}
\end{figure}

In order to understand the nature of emitting sources, we measure the spectral energy distributions (SEDs) within several apertures across SN\,1987A.
The locations of apertures are marked in Fig.\,\ref{fig:SED-region}, and the extracted SEDs are plotted in Fig.\,\ref{fig:SED}.
The `background' or ISM continuum level, which is estimated from the nearby field (Sect.~\ref{sect-observations}), has been subtracted beforehand.

The fluxes at F323N, F356W and F444W follow a power-law continuum across SN\,1987A (black lines in Fig\,\ref{fig:SED}).
The ejecta have the steepest  power-law index $\alpha$ 
at a value of 3.1, where $\alpha$ is defined as $F_{\nu} \propto \nu^{-\alpha}$.
Apart from the ejecta,
as expected from the \Fviii/\Fvi\, ratio map in Fig.~\ref{fig:color}, the steepest power-law index is found on the western side of the outer spots. 
The power-law indices are $\alpha=2.9$ and 1.9 at the western and eastern outer spots, respectively. 
The power-law indices are as small as $\alpha=1.6$ at the hot spots on the ring and $\alpha=1.5$ within the diffuse emission.

\subsubsection{NIR continuum: synchrotron and dust}

\begin{figure}
	 \includegraphics[height=6.cm, trim={0cm 0cm 0cm 0cm},clip]{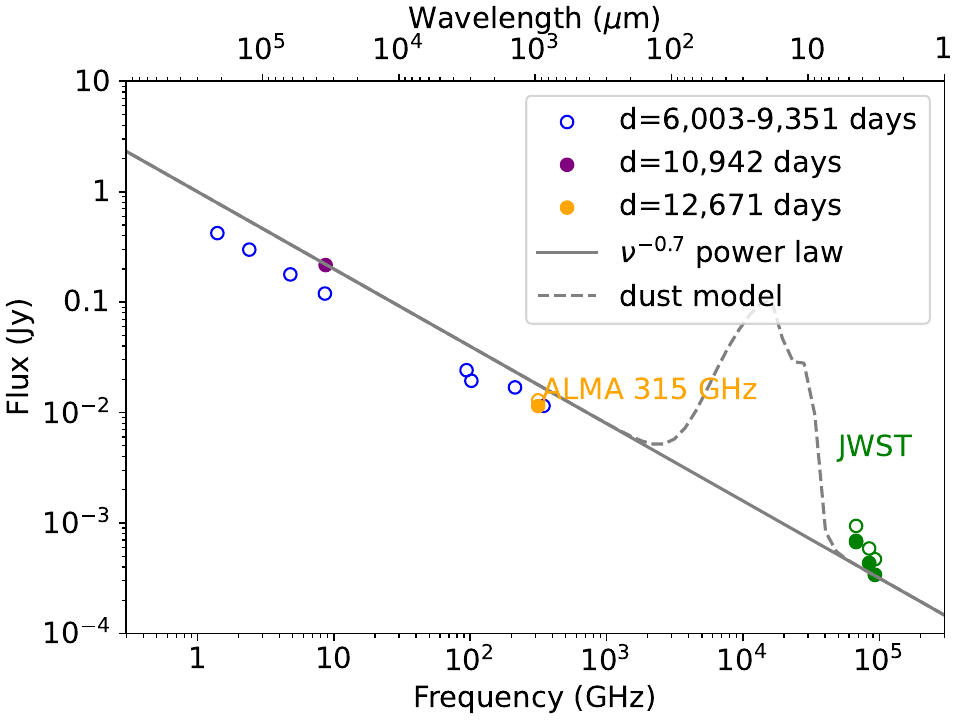}
        \caption{ The spectral energy distributions from near-infrared to radio in the past two years and the past. The filled circles of ALMA and {\it JWST}/NIRCam data points show the fluxes of the equatorial ring and the outer spots.    For  {\it JWST}/NIRCam, additionally the diffuse emission is included in the fluxes plotted with empty circles.  The grey line shows the `warm' dust model from  \citet{Matsuura.2022fk} 
        \label{fig:radio}}
\end{figure}

In light of the power-law SEDs in the 3--4\,$\mu$m wavelength range, we discuss the sources of this emission in the equatorial ring and its exterior.
Historically, the emission at 3--4\,$\mu$m was solely attributed to dust thermal emission from the equatorial ring, but synchrotron emission can be an additional and important source at the current epoch.
In shocked gas, fast-moving electrons collide with dust grains and collisionally heated dust grains, which have a temperature of 180--200\,K  \citep{Dwek:2010kv, Matsuura.20185p8}, and radiate their energy at the MIR wavelengths (Fig.\ref{fig:radio}). 
The modified blackbody emission from the warm dust drops sharply from the SED peak towards NIR wavelengths, in accordance with the Wien's law approximation to the Planck function.
Thus, NIR emission has been 
explained by additional `hot' \citep[370--460\,K; ][]{Dwek:2010kv} dust consisting of smaller grains.
Because the heat capacities of small grains are lower, they can reach much higher energy than the larger grains.

On the other hand, our total SED analysis suggests that synchrotron emission can contribute substantially to the NIR flux in the ring and its exterior.
Fig.~\ref{fig:radio} shows the integrated SED of the ring and outer hot spots from the IR to the millimetre wavelengths.
The total fluxes in the {\it JWST} F323N, F356W and F444W bands are plotted in open green circles, where the aperture sizes are described in Appendix.\,\ref{app-flux-measurements}.
The historical measurements of the radio synchrotron emission before 2014 are plotted in blue circles, and they demonstrate the power-law of the synchrotron emission \citep{Zanardo:2010p29272, Zanardo:2013fn, Zanardo:2014gu}. The day 10,942 flux measurement at 8.7\,GHz from \citet{Cendes.2018z98} is plotted with the filled purple circle.
The total ALMA 315 GHz flux in 2021 is plotted with an orange open circle and the flux excluding the ejecta is plotted in a filled orange circle.
 The radio fluxes continued to increase since 2014 until at least until 2017 \citep{Cendes.2018z98}, so that pre-2014 fluxes are lower than the latest (day=10,942) flux measurement at 8.7\,GHz \citep{Cendes.2018z98}.
The power-law index at the millimetre and submillimeter wavelengths has been estimated as $\alpha = 0.74$ by \cite{Zanardo:2013fn}, fitting the blue open circles. More recently, \citet{Cigan:2019cl} estimated $\alpha=0.70$.

In Fig.~\ref{fig:radio}, the grey line shows the power-law of $\alpha=0.70$ crossing the 8.7\,GHz flux at day=10,942.
Even though it is not intended to fit the JWST fluxes, the grey line can explain the F323N and F356W fluxes very well, 
showing that a substantial fraction of the NIR fluxes is synchrotron emission.  
The ALMA flux at 315\,GHz is probably underestimated, because the interferometry had a fully recoverable size of only $0.99''$, and any structure extending beyond that size, such as diffuse emission, is over-resolved and its flux is not properly accounted.

Similar analysis of the {\it JWST} NIRSpec and MIRI/MRS total SED also found that synchrotron radiation is the most important source of the continuum in the 3--4\,$\mu$m range \citep{Jones.20232sn}. 
Hot  ($\sim$300\,K) dust starts to contribute to F444W, only.

In other young SNRs, Cassiopeia A and the Crab Nebula, the synchrotron emission contributes to the continuum emission from infrared wavelengths at 2~$\mu$m to cm wavelengths \citep{2003ApJ...587..227J, Gomez:2012fm, 2021MNRAS.502.1026D}.
It is not surprising synchrotron emission contributes to the continuum in SN\,1987A, too.

\citeauthor{Jones.20232sn} showed that the total SED of SN 1987A at 3--4\,$\mu$m additionally contains minor contributions from free-free and bound-free emission.
However, their flux density decreases at longer wavelengths, which is opposite to the increasing continuum with longer wavelengths in the measured spectra.
Hence, the contributions of free-free and bound-free emissions into the  3--4\,$\mu$m spectra are likely very minor.

\subsubsection{Spatial distributions of dust and synchrotron}

Now, we carefully examine the spatial distributions of dust and synchrotron emission.


The F356W image is compared with the ALMA continuum image in Fig.~\ref{fig:comparison_images_ALMA} right panel.
The ALMA continuum image traces synchrotron in the ring and its exterior, though it does trace dust in the ejecta.
In 2021, the emitting regions of the ALMA 315\,GHz  (950\,\micron)  flux 
correspond to hot spots within the equatorial ring and the outer spots exterior to the equatorial ring (Fig.~\ref{fig:comparison_images_ALMA}). In particular, on the eastern side, the 315\,GHz emission is found mostly in the outer spots, and hardly found in the hot spots.

\begin{figure*}
	 \includegraphics[width=8.8cm, trim={0.4cm 0.4cm 0.4cm 0.4cm},clip]{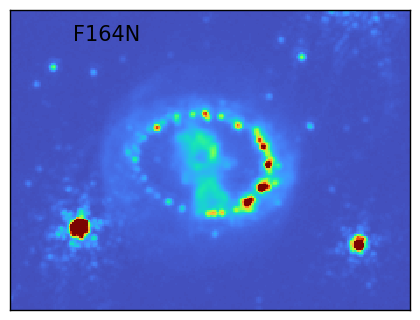}
	 \includegraphics[width=8.8cm, trim={0.4cm 0.4cm 0.4cm 0.4cm},clip]{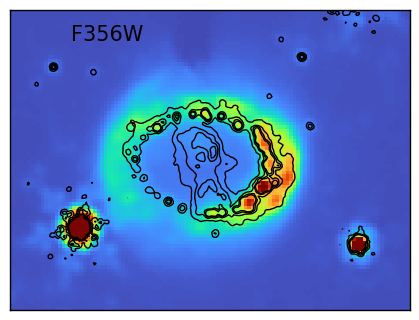}
	 \includegraphics[width=8.8cm, trim={0.4cm 0.4cm 0.4cm 0.4cm},clip]{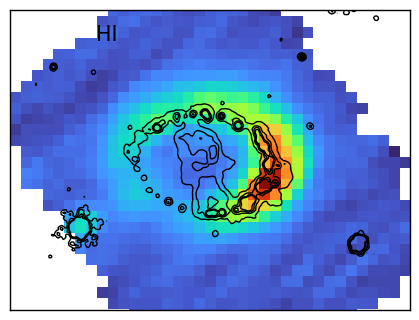}
	 \includegraphics[width=8.8cm, trim={0.4cm 0.4cm 0.4cm 0.4cm},clip]{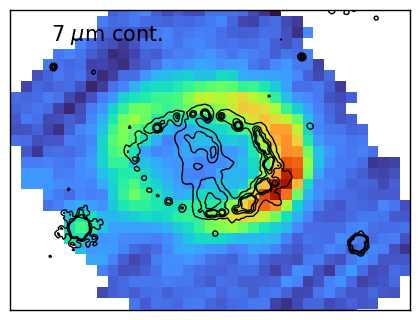}
        \caption{ Comparison of the emitting regions of ionised gas (F164N and 7.46\,\micron\,\HI\,6--5 Pf$\alpha$ lines), F356W continuum and 7\,$\mu$m (7.43--7.45\,\micron) continuum. The  7.46\,\micron\,\HI\,6--5 Pf$\alpha$ lines and 7\,$\mu$m continuum images are from MIRI/MRS, replicated from \citet{Jones.20232sn}. The black contour lines trace F164N, though due to slightly different smoothing of the F164N data to the new pixel scales, there is a subtle difference in contour lines from one image to the other. The dust continuum, represented by the 7\,$\mu$m continuum, is mainly emitted from outer spots and diffuse emission, while F356W emission is found in both hot spots and outer spots, as well as diffuse emission.
        \label{fig:dust_images}}
\end{figure*}

Fig.~\ref{fig:dust_images} compares the F356W image and dust continuum image.
The left top panel of the figure shows the F164N image, and contours of the F164N image are plotted in the other three panels in order to guide the locations of hot spots and diffuse emission.
The lower panels are MIRI/MRS images, replicated from \citet{Jones.20232sn}. 
The left lower panel shows  \HI\, 7.46\,$\mu$m image, demonstrating that ionised gas is emitted from the hot spots but only weakly from the outer spots, similar to F164N ring image.
In contrast, the 7.4436--7.4572\,$\mu$m dust image (the right lower panel) reveals that the dust emission is from the exterior to the equatorial ring, i.e. either outer spots or diffuse component.
These outer spots and diffuse emission have the steepest power-law index at F444W/F356W (Fig.\ref{fig:color}).
The dust thermal emission at $\lesssim 7$ $\mu$m corresponds to the Wien's law regime of the blackbody, which 
can explain a steep ($\alpha\sim 3$; Fig.\,\ref{fig:SED}) power law index. 
The brightest hot spots within the western ring in F356W image (right top panel) have little dust emission at the 7 $\mu$m continuum (right bottom panel), hence, these hot spots are most likely synchrotron-dominated.

There is a morphological difference amongst dust, synchrotron, and atomic lines (\Ha\, and \FeII):
dust emission appears at the limited locations of the diffuse emission and outer spots, while synchrotron emission is found in both outer spots and hot spots, as well as diffuse emission. \Ha\, and \FeII emission is strongest at the hot spots.
The difference in emitting locations is caused by the timescale of these three physical processes.
More narrowly confined dust emitting regions in comparison with synchrotron emitting regions is explained by the shorter cooling time scale of dust grains than synchrotron electrons in the post-shocked region.
The radially expanding shocks gradually engulf new circumstellar material, including dust grains. 
Fast-moving electrons in hot shocked gas collide with dust grains, heating these grains.
The heated grains emit the NIR and MIR and radiation, which also cools down dust grains and gas in post-shocked regions.
Further, a high collision of charged particles can sputter atoms from the surface of dust grains, and making dust grain size smaller (dust destruction).
The cooling time scale was predicted to be 12--22 years at a gas density of $n_{\rm H}=10^4$~cm$^{-3}$ for grain sizes of $>$0.2~$\mu$m, and at a post-shock velocity of 600--700 km\,s$^{-1}$.  Smaller grains cool faster than that. Whereas the dust destruction time scale was predicted to be 4--15 years \citep{Dwek:2010kv}.
Overall the shocks have passed the equatorial ring after about 2010 \citep{Fransson:2015gp}, although the western side of the ring seems to have interacted until about 2014 \citep{Larsson.2019h9}.
Little dust emission from the hot spots within the equatorial ring in 2022 shows that either the cooling timescale or destruction timescale of dust grains is less than 8 years.
If the cooling rate is responsible for fading MIR dust emission, the grain size can be smaller than 0.2~$\mu$m.
Otherwise, dust grains in the hot spots are destroyed, hence, little emission is left at the hot spots.

On the other hand, the radiative loss timescale of relativistic electons may be longer than the ten-year timescale, as discussed in the next section, hence, synchrotron radiation is still emitted in hot spots, even if the shocks have passed 12 years ago.

 The brightness of the hydrogen recombination lines also changes within the recombination time scale, which 
 is $\tau=(\alpha_B n_{\rm H})^{-1}=1.22\times10^5 / n_{\rm H}$\,years \citep[Eq. 15.7 of ][]{2011piim.book.....D}, where $n_{\rm H}$ is the hydrogen number density in cm$^{-3}$.
With estimated density of $n_{\rm H}=10^4$~cm$^{-3}$, the recombination time scale is 12 years. Having \Ha\, in the western hot spots still but gradually fading since 2014 matches with explanation by this recombination timescale.

\subsubsection{Broken power-law synchrotron emission at NIR wavelengths}

Synchrotron radiation makes a substantial contribution to the NIR continuum.
The NIR power law index of the continuum is about $\alpha$=1.5--1.6 at hot spots.
This is higher than the millimetre power law index, measured to be  $\alpha$=0.7--0.74 \citep{Zanardo:2013fn, Cigan:2019cl}.
Having little MIR dust emission at hot spots, it is very unlikely that the tail of dust emission increases the power law at hot spots.
The synchrotron power law index at NIR wavelengths is likely to be intrinsically higher than that at millimetre wavelengths at hot spots.

Though synchrotron radiation, in general, follows a power-law ($F_\nu\propto\nu^{-\alpha}$), 
the radiative lifetime of the more energetic electrons, which are responsible of the emission at higher frequencies, is shorter, leading to a steepening of the spectrum at higher frequencies \citep{1965ARA&A...3..297G, 2011hea..book.....L, 2021MNRAS.502.1026D}.
 The breakpoint of the power-law index is in general found at optical and mid-infrared wavelengths in SN remnants \citep{Buhler:2014fo, 2021MNRAS.502.1026D}.
Having a broken law in the IR wavelengths may explain the steeper NIRCam power-law index than that in the millimetre wavelength.

Moreover, previous millimetre power-law index measurements suggest the synchrotron power-law index itself could have some spatial variations; $\alpha=0.6$--$1.0$ with an aperture of 0.8''
\citep{Zanardo:2014gu}.

We test if such a broken IR law of the synchrotron emission is feasible in SN\,1987A.
If a relativistic electron can lose energy by synchrotron radiation before an electron accelerates to that energy (or frequency), there is no electron emitting synchrotron radiation at that frequency. The typical frequency, $\nu_{\rm br}$, where radiative loss affects the spectrum is given by
 \begin{equation} 
	\nu_{\rm br} \approx 7.2 \times 10^{16}  \left( \frac{ B }{100 \mu {\rm G}} \right)^{-3}  \left(  \frac{ t_{\rm age}} {100 {\rm yr}} \right)^{-2} {\rm Hz}
\end{equation} 
where $B$ is the magnetic field, and $ t_{\rm age}$ is the electron's  radiative lifetime
 \citep{2021MNRAS.502.1026D} \citep[also in][]{1970ranp.book.....P}.
 The collision of the blast wave with the equatorial ring occurred somewhere between 1999 and 2005
 \citep{Luo.1991, Chevalier.1995, McCray:2016bg}, so that the lifetime should be less than 27\,years.
Having the break wavelength/frequency to be at 3\,\micron\,  requires the magnetic field of $B\ge2.1$\,mG. 
The polarization measurement of the millimetre synchrotron emission estimated the magnetic field of a few mG at the post-shocked region in 2015--2016 \citep{Zanardo:ty}.
This magnetic field can cause a broken power law index of the synchrotron emission.

Theoretical models of SN shock evolution show a broken power-law in the presence of a strong magnetic field.
 \citet{Berezhko.2011}  predicted the broken law to be present at IR wavelengths in SN\,1987A.
The model involved a 20\,mG magnetic field downstream, as this could explain non-detection of synchrotron emission in X-ray in their model.
In this case, the synchrotron broken law already appears at the MIR, and continues with the steep power-law index into the NIR.
 The predicted spectra had time variation, and the IR power-law index of about 1.2 was predicted in 2020.
 This predicted index is slightly smaller than the observed one, but much steeper than the millimetre power-law index.
On the other hand, \citet{Petruk.202280q} predicted a much lower magnetic field ($<20\mu$G), and in this case, the power-law break falls at a shorter wavelength than the NIR.
These contrasting predictions show that the measurement of a magnetic field is the key next step.

\subsubsection{The NIR continuum as a tracer of shocks and ionised lines as a tracer of recombining regions }

Historically, synchrotron emission, detected in radio and X-ray wavelengths, was found only from the equatorial ring, and the outer spots were not considered \citep[e.g.][]{2002ApJ...567..314P, Ng:2013bt, Zanardo:2014gu}. 
That was definitely the case at least until about 2014, when the blast waves were about to pass the equatorial ring \citep{Fransson:2015gp}.
Although not fully resolved down to the scale of the hot spots and outer spots, the radii of the radio and the X-ray emitting ring sizes have continued to increase since then \citep{Frank:2016ka, Cendes.2018z98}. This trend is more prominent in the hard (0.5--10 k eV) X-ray emission.
The soft (0.3--0.8\,keV) X-ray emission radius stagnated around day 6000 in 2004 \citep{Frank:2016ka}, and it is considered to be from hot spots in the equatorial ring \citep{2012ApJ...752..103D}.
In the ALMA image taken in 2015, the synchrotron emission extended just beyond the equatorial ring \citep{Cigan:2019cl}.

Now, the synchrotron emission in X-ray and radio wavelengths must be emitted by both the ring hot spots and the outer spots, as shown by the {\it JWST} F323N, F356W, and F444W images and the ALMA 315~GHz image. 
That would explain the recent increase in the radii of the radio and hard X-ray emission.
Considering that the brightness from the west side of the equatorial ring and outer spots dominates the overall brightness in the F323N, F356W, F444W, and radio flux, as well as X-ray, the current interactions of shocks with the circumstellar material must be dominant on the western side of the equatorial ring and the material beyond the ring. 
The majority of the blast waves have already passed the equatorial ring on the eastern side.


Shocked gas is typically traced by X-rays, and the electron temperature is estimated to be 1--3$\times10^7$ K \citep{Ravi.2021}. 
Around that temperature, in case A, hydrogen recombination lines have emissivity decreasing with increasing temperature, as $T^{-0.92}$ \citep{Storey.1995}.
The outer spots, where current interaction is ongoing, are expected to have a higher temperature than the `hot spots'
 (Note that the name of `hot spots' is based on observations in 1990s and the temperature has changed since, and does not reflect the current temperature.).
This explains why hot spots are brighter than the outer spots in hydrogen recombination lines, such as \Bra\,(\Fvii\, in Fig.\ref{fig:long_images})
and {\it HST} \Ha\, images \citep{Larsson.2019h9}.
That makes the contrast between the hydrogen recombination line images, which show the hot spots to be brighter than the outer spots, and the continuum images ({\it JWST} F323N, F356W, and F444W images), which have equally bright hot spots and outer spots.

The continuum may contain some emission from small dust grains \citep{Dwek:2008p28793, Matsuura.2022fk, 2023arXiv230913011A, Jones.20232sn}.  The emission is from collisionally heated dust in shocks.
Even if arising from a combination of synchrotron and dust emission, the power-law continuum emission in the NIR wavelength would be a good tracer of the shocked region.

In contrast, the ionised lines originate from hot spots within the equatorial ring, which contain more substantial gas mass but have lower temperatures. 
The recombination timescale is much longer than the synchrotron cooling time, and a lower temperature favours higher recombination line emissivity. The hot spots within the equatorial ring thus remain brighter for longer timescales in recombination lines.

\subsection{Outer rings} \label{sec-outer-ring}

The outer rings are considered to have been expelled from the progenitor about 20,000 years before the supernova explosion \citep{1996ApJ...459L..17P, 2000ApJ...528..426C}. 
The outer rings have been detected mainly in atomic lines, such as \NII, {[\ion{O}{iii}]} and H$\alpha$ in the optical wavelengths \citep{1996ApJ...459L..17P, Tziamtzis.2010}, and Ne and  {[\ion{O}{i}]} lines in the MIR wavelengths \citep{Jones.20232sn}.

\begin{figure*}
	 \includegraphics[width=0.47\textwidth, trim={0.1cm 0.2cm 2.7cm 0.6cm},clip]{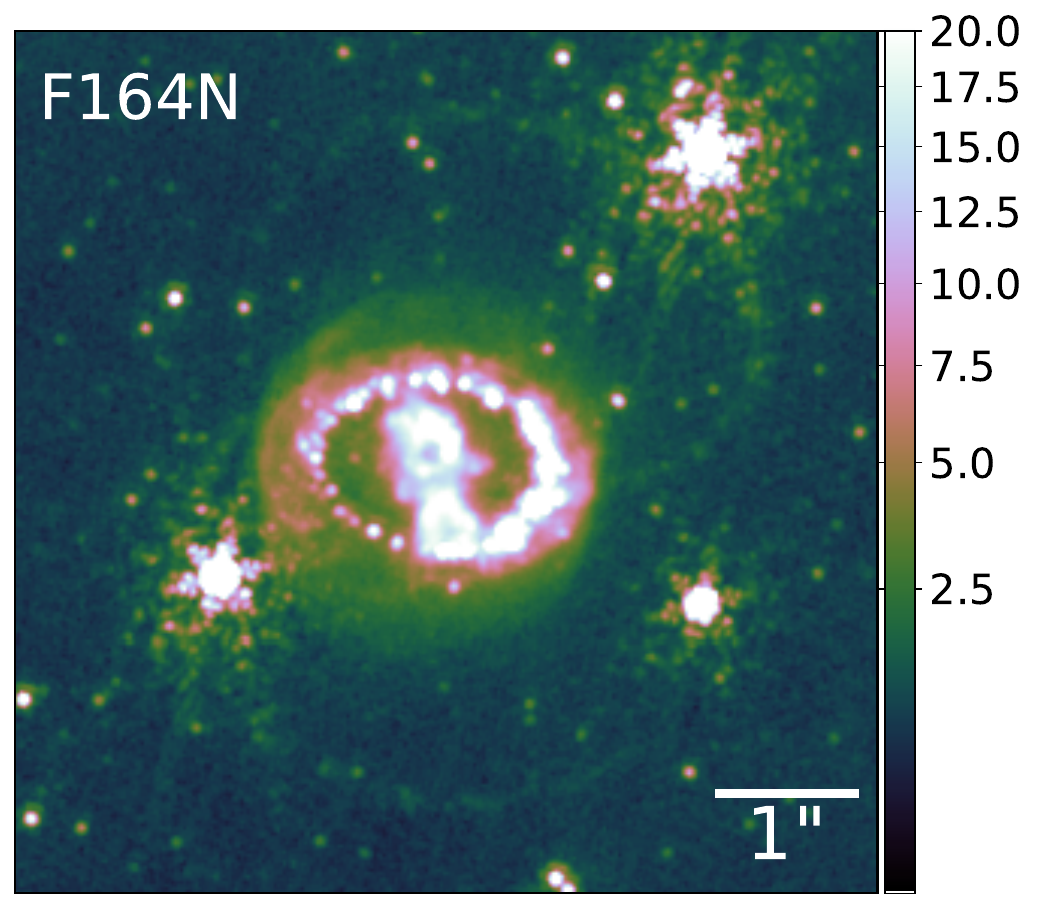}
	 \includegraphics[width=0.47\textwidth, trim={0.1cm 0.2cm 2.05cm 0.6cm},clip]{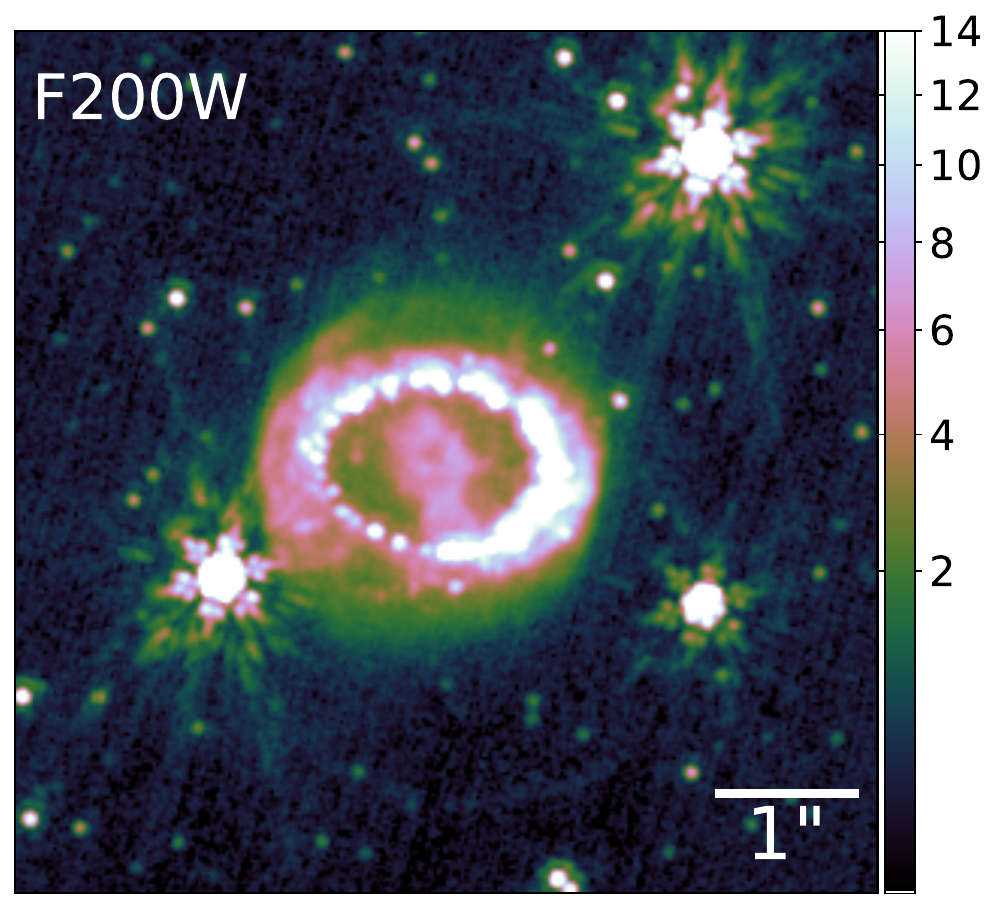}
   \includegraphics[width=0.47\textwidth, trim={8.6cm 2.0cm 10.6cm 2.3cm},clip]{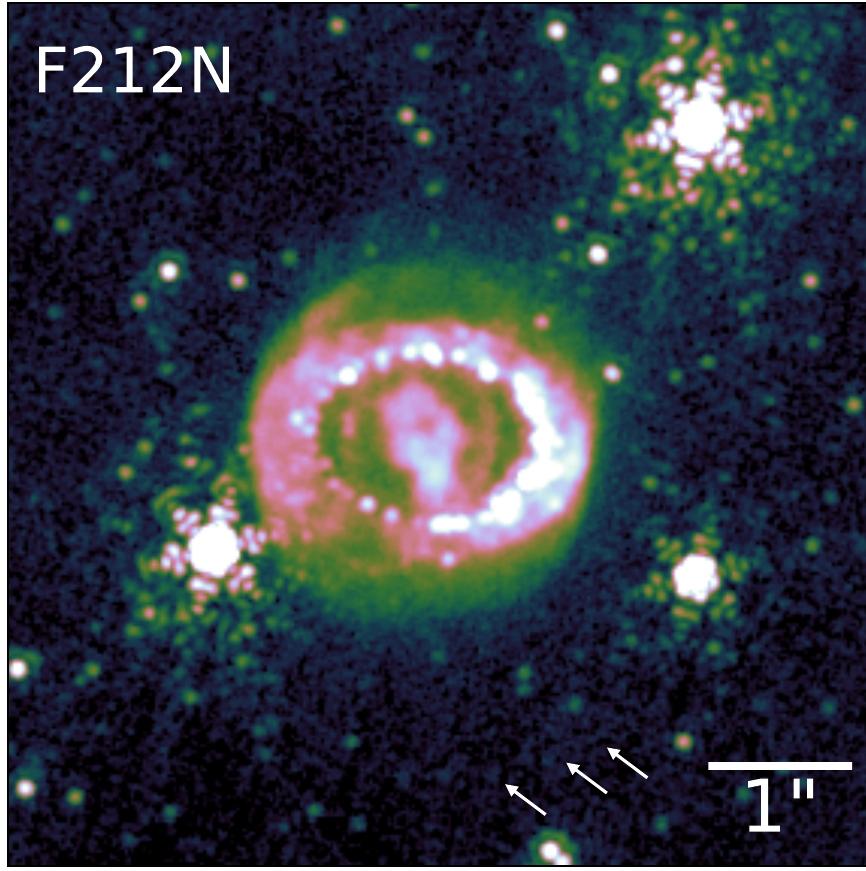}
	 \includegraphics[width=0.47\textwidth, trim={0.1cm 0.2cm 2.05cm 0.6cm},clip]{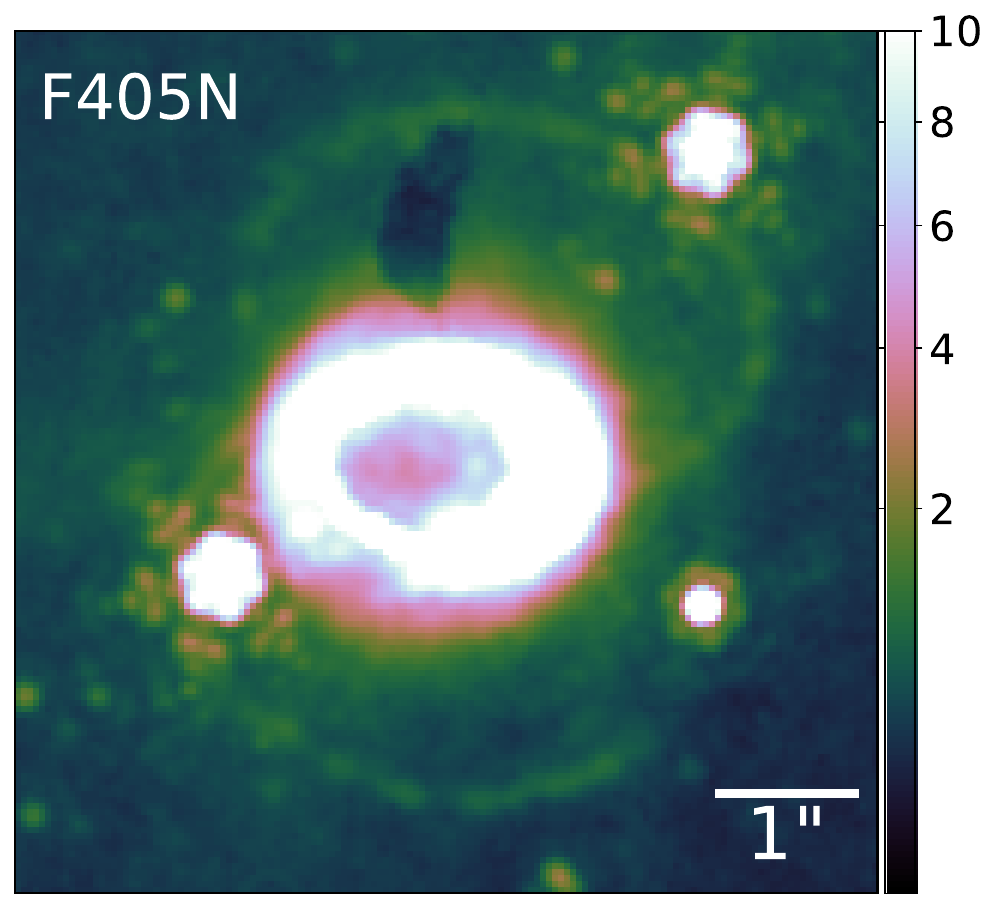}
        \caption{The detection of the outer rings in four filter bands. The arrows in the F212N image show the locations of potential detections of the outer rings. The black shadow in the F405N is an artefact, due to incomplete flat-field correction. 
        \label{fig:outer_ring}}
\end{figure*}

In our NIRCam images, very faint emission from the outer rings was detected in three filter bands F164N, F200W and F405N (Fig.\,\ref{fig:outer_ring}). 
All three filter bandpasses contain atomic lines: F164N contains 1.641\,$\mu$m and \FeII\, line, F200W  is mainly from  1.875\,$\mu$m \Paa, and 2.059\,$\mu$m \HeI\, lines, and F405N is from \Bra.
A hint of part of the outer rings is recognisable in the F212N image, too.
Non-detections in the continuum-dominated filters may suggest that the emission from the outer rings is dominated by the atomic lines with very faint H$_2$ in F212N.
This supports the proposition that the gas in the outer rings was ionised by the initial supernova flash \citep{Tziamtzis.2010} and that lines from the ionised gas can be still detected at this late epoch.

The outer rings are known to cross the ejecta and the equatorial ring as seen in projection from Earth \citep{Tziamtzis.2010}. 
We assess if the emission from the outer rings may still contribute to some structures in the {\it JWST} image in 2022 in Sect.\,\ref{sec-outerring}. 
From the measurement of the surface brightness,
the outer ring emission accounts for a negligible fraction (less than 10\,\%) of the surface brightness of the ejecta in F164N and F200W images.

\section{Conclusions}

Very deep and high-angular resolution NIRCam imaging of Supernova 1987A reveals unprecedented details of near-infrared emission from this young supernova remnant. 
This includes the detection of the NIR synchrotron emission, the discovery of the crescents, and the characterisation of the bar, which is a substructure of the ejecta.

Regions currently undergoing shocks are traced by the NIR continuum through synchrotron and thermal dust emission at 3--5\,\micron. This continuum emission is spatially resolved in NIRCam images, and is found in hot spots within the equatorial ring, in outer spots exterior to the equatorial ring, as well as in the diffuse emission exterior to the equatorial ring. 
At 3--5\,\micron, the outer spots are as bright as the hot spots,  despite the fact that the hot spots have the highest density within the circumstellar system. This is probably because stronger magnetic fields and enhanced local density at shocks increase the synchrotron radiation in the outer spots. Future monitoring and modelling of synchrotron emission will reveal how the synchrotron emission evolves with time in an SNR, after the passage of the shock front. 
Such time sequence observations will provide a rare opportunity amongst SNRs that would constrain hydrodynamic models of SN shocks \citep[e.g.][]{Borkowski.1997, 2012ApJ...752..103D, Kirchschlager.2019yrq, 2020A&A...636A..22O}, and particle acceleration in the shocked regions \citep[e.g.][]{2021MNRAS.502.1026D}.
With the advent of high sensitivity and high angular resolution images provided by {\it JWST}/NIRCam, our observations of SN\,1987A demonstrate that NIRCam opens up a window to study particle-acceleration physics probed by NIR synchrotron emission in supernova remnants and other synchrotron emitting objects such as active galactic nuclei and blazars.

The NIRCam 1--2\,$\mu$m images reveal the clumpy structure of the ejecta, including the bar. The presence of clumpy structure is consistent with SN explosion models that predict the fragmentation of the SN ejecta into clumps due to Rayleigh-Taylor instabilities, which are triggered by the explosions  \citep{Wongwathanarat:2015jv}.

The NIRCam images show that the centre of the ejecta is still obscured by dust, and seen as a `hole' at the 3--5\,\micron\, wavelengths.
The presence of dust obscuration in the ejecta means that optical and near-infrared emission mainly traces the foreground of this heavily self-absorbed dusty region.  
As this heavily self-absorbed dust region expands further, the optical depth of the dust will decrease in time. 
Once the dust optical depth has dropped sufficiently, a complete picture of the ejecta will be visible at the NIR wavelengths. 
A new feature, the crescents, is discovered between the ejecta and the equatorial ring. 
 The crescents are probably either the ionisation front of the inner ejecta, irradiated by X-ray and UV emitted from the outer spots,
or tthe edge of the reverse shocks bounced back from the equatorial ring.
Details of our analysis of these crescents will be published elsewhere.


Very deep and high angular resolution NIRCam images revealed detailed structures in the young and close SNR, SN\,1987A. They will be used to understand the physical processes that shape complex structures of SNe. The processes to be investigated include mass loss from the progenitor star, the SN explosion, dynamical processes, heating (by X-ray and UV radiation, and $^{44}$Ti decay, and potentially the neutron star) and cooling processes (atomic and dust radiations, as well as adiabatic). All of these processes are imprinted into the complexity of emissions from the ejecta and circumstellar matter in this SNR. The next stage is to disentangle all of these processes to truly understand the cause of the complexity and time evolution of the SNRs.

\section*{Acknowledgements}
This work is based on observations made with the NASA/ESA/CSA James Webb Space Telescope. The data were obtained from the Mikulski Archive for Space Telescopes at the Space Telescope Science Institute, which is operated by the Association of Universities for Research in Astronomy, Inc., under NASA contract NAS 5-03127 for JWST. These observations are associated with program \#1726.

This paper makes use of the following ALMA data: ADS/JAO.ALMA\#2021.1.00707.S. ALMA is a partnership of ESO (representing its member states), NSF (USA) and NINS (Japan), together with NRC (Canada), MOST and ASIAA (Taiwan), and KASI (Republic of Korea), in cooperation with the Republic of Chile. The Joint ALMA Observatory is operated by ESO, AUI/NRAO and NAOJ.

This work also utilises JWST data from the program \# 1232, obtained from the Mikulski Archive.

This work presents results from the European Space Agency (ESA) space mission Gaia. Gaia data are being processed by the Gaia Data Processing and Analysis Consortium (DPAC). Funding for the DPAC is provided by national institutions, in particular the institutions participating in the Gaia MultiLateral Agreement (MLA).

 M.M. and R.W. acknowledge support from the STFC Consolidated grant (ST/W000830/1).
M.J.B., and R.W. acknowledge support from European Research Council (ERC) Advanced Grant SNDUST 694520. I.D.L. and F.K. acknowledge funding from the European Research Council (ERC) under the European Union's Horizon 2020 research and innovation programme (\#851622 DustOrigin). RDG was supported, in part, by the United States Air Force.
Work by R.G.A. was supported by NASA under award number 80GSFC21M0002.
APR and SP are supported in part by the STScI grant, JWST-GO-01726.032-A
C.G. is supported by a VILLUM FONDEN Young Investigator Grant (project number 25501).

\section*{Data availability}




The pipeline reduced data are available at MIKULSKI archive 
(\url{http://dx.doi.org/10.17909/dzkq-7c90}).

\bibliography{sn1987a_jwst}
\bibliographystyle{mn2e}


\appendix

\section{Contribution of the Outer Ring onto the ejecta}  \label{sec-outerring}

\begin{figure*}
	 \includegraphics[height=7.8cm, trim={2.5cm 3cm 2.5cm 3.6cm},clip]{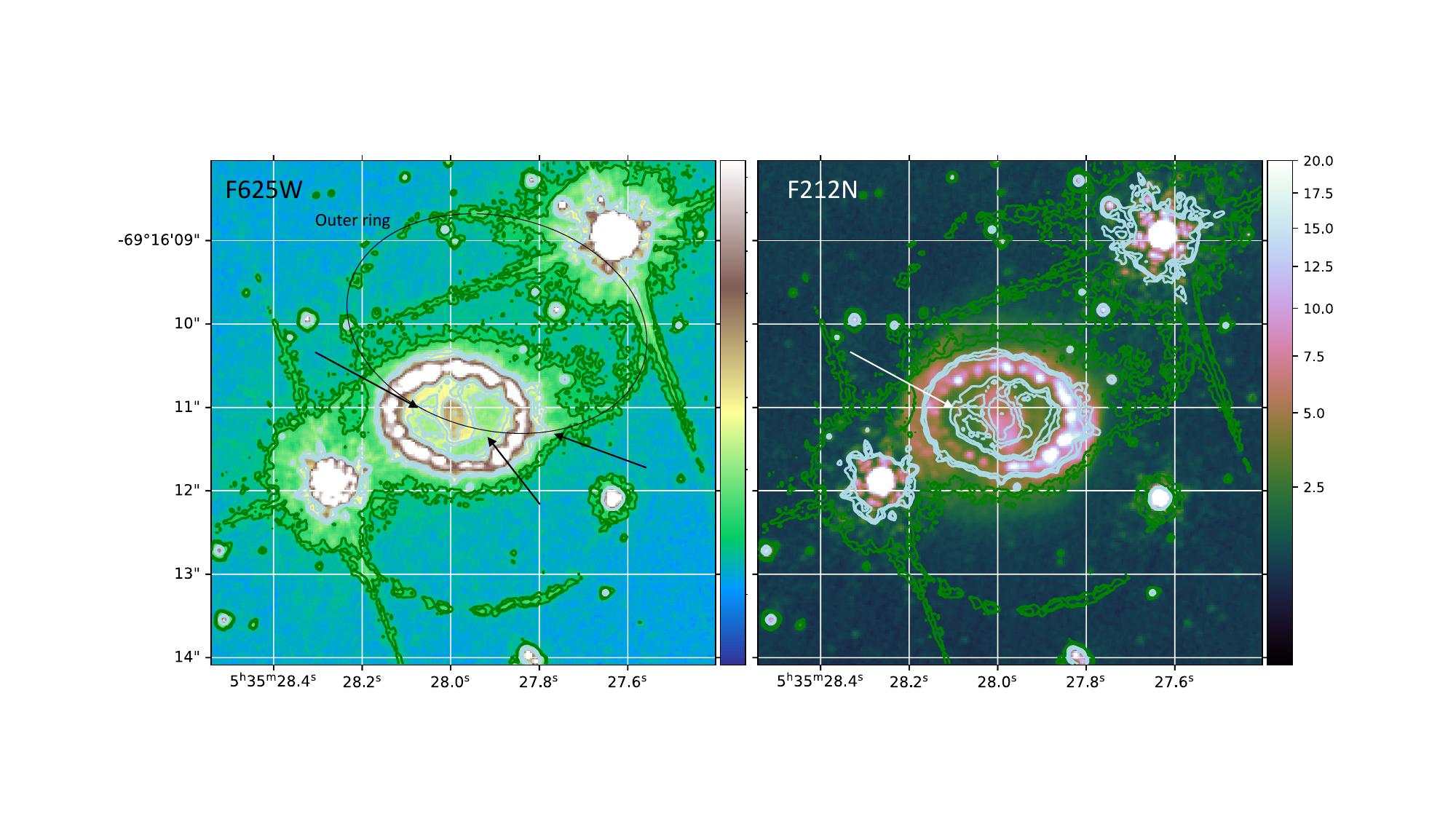}
        \caption{({\it Left:}) HST F625W image in 2003 \citep{Tziamtzis.2010}, showing the outer rings in the line contours. 
        The north outer ring crosses the equatorial ring and ejecta, indicated by three black arrows. Since then, the outer ring has faded gradually.
        {\it(Right:)} F212N image in colour contours, overlayed with line contours identical to the left panel (green and white). The colour bar is in the unit of MJy~sr$^{-1}$.
         The local peak on the east crescent could have some gain of the surface brightness from the outer rings. 
        \label{fig:comparison_images_outerring}}
\end{figure*}

The outer rings are known to cross the ejecta and the equatorial ring as seen in projection from Earth \citep{Tziamtzis.2010}. 
We assess if the emission from the outer rings may still contribute to some structures in the {\it JWST} image in 2022. 
The outer rings have been gradually fading since the {\it HST} image was taken in 2003  \citep{Tziamtzis.2010}, 
and we use the 2003 {\it HST} image as a guide of the outer ring, rather than more recent {\it HST} images.
The left panel of Fig.~\ref{fig:comparison_images_outerring} shows the {\it HST} F625W image. The levels of the line contours are adjusted to highlight the outer rings, the equatorial ring and the ejecta. The northern outer ring is encircled with a black line. This ring is highlighted at three locations using black arrows: the ring enters the equatorial ring from the north-east, and proceeds towards the centre of the equatorial ring, then exits from the south-west of the ejecta. Afterwards, it crosses the equatorial ring for the second time. Grey contour lines show some local increase in intensities at these three locations.
The right panel of Fig.~\ref{fig:comparison_images_outerring} shows the NIRCam F212N image of the same region in the colour contour. The line contours are taken from the {\it HST} F625W image. At the position of the white arrow, the north outer ring and the east crescent meet. This crossing point coincides with the local peak of the crescent. This local peak might have some gain in the intensity from the outer ring.
Alternatively, the colour seems to be blue, which might be a sign of a field star at this location.
From the measurement of the surface brightness,
the outer ring emission accounts for a negligible fraction (less than 10\,\%) of the surface brightness of the ejecta in F164N and F200W images.

\section{Measurements of the fluxes in the equatorial ring and outer spots} \label{app-flux-measurements}

In order to measure the NIR continuum fluxes in the equatorial ring and outer spots, two different types of elliptical apertures were used.
These fluxes were measured with an elliptical aperture of  1.50'' and 1.37'' major and minor axes (Fig.\,\ref{fig:SED-region}), with a negligible contribution from field stars within the aperture.
The filled circles show the fluxes measured from another elliptical annulus that includes the ring and the outer spots, but excludes the ejecta and diffuse emission outside the outer spots.
The major axes of the inner and outer annulus are 0.65" and 1.28" and minor axes of these are 0.52" and 0.90". 





\bsp	
\label{lastpage}
\end{document}